\documentclass[structabstract]{aa}  
\usepackage[dvips]{graphicx}
\usepackage[comma,authoryear]{natbib}
\usepackage[english]{babel}
\usepackage[latin1]{inputenc}
\usepackage{amssymb}
\usepackage{hyperref}
\bibpunct{(}{)}{,}{a}{}{,}%

\newlength{\mycolumn}
\newcommand{\URL}[1]{\href{#1}{\texttt{#1}}}

\renewcommand{\l}{\ell}
\renewcommand{\d}{\mathrm{d}}
\newcommand{\ie}{\textit{i.e.}}

\newcommand{\eg}{\textit{e.g.}}
\newcommand{\Kavg}{K_{\mathrm{avg}}}
\newcommand{\Kcross}{K_{\mathrm{cross}}}
\newcommand{\rhoref}{\rho_{\mathrm{R}}}
\newcommand{\Teff}{T_{\mathrm{eff}}}
\newcommand{\logg}{\log(g)}
\newcommand{\smax}{s_{\mathrm{max}}}
\newcommand{\rhomax}{\bar{\rho}_{\mathrm{max}}}
\newcommand{\Msurf}{M_{\mathrm{surf}}}
\newcommand{\Aprime}{A$^{\prime}$}
\newcommand{\rhobar}{\bar{\rho}}
\newcommand{\gcm}{\mathrm{g}\,\mathrm{cm}^{-3}}

\begin{document}

\title{Estimating stellar mean density through seismic inversions}

\author{D. R. Reese\inst{1,2}
        \and
        J. P. Marques\inst{1,3}
        \and
        M. J. Goupil\inst{1}
        \and
        M. J. Thompson\inst{4,5}
        \and
        S. Deheuvels\inst{6}
       }
% LESIA, Observatoire de Paris, CNRS, UPMC, Université Paris-Diderot, 5 place Jules Janssen, 92195 Meudon, France

\institute{LESIA -- Observatoire de Paris, CNRS, UPMC Univ. Paris 06,
           Univ. Paris-Diderot, 5 place Jules Janssen, 92195 Meudon, France
          \and
           Institut d'Astrophysique et Géophysique de l'Université de Liège,
           Allée du 6 Août 17, 4000 Liège, Belgium \\
           \email{daniel.reese@ulg.ac.be}
          \and
          Georg-August-Universität Göttingen, Institut für Astrophysik, 37077 Göttingen, Germany
          \and
           High Altitude Observatory, National Center for Atmospheric Research, Boulder, CO 80307, USA
           \and
           The School of Mathematics and Statistics, University of Sheffield, Hicks
           Building, Hounsfield Road, S3 7RH, Sheffield, UK
           \and
           Department of Astronomy, Yale University, P.O. Box 208101, New Haven, CT 06520-8101, USA
          }

\date{}

\abstract
% context heading (optional)
{Determining the mass of stars is crucial both to improving stellar evolution
theory and to characterising exoplanetary systems.  Asteroseismology offers a
promising way to estimate stellar mean density.  When combined with accurate
radii determinations, such as is expected from GAIA, this yields accurate
stellar masses.  The main difficulty is finding the best way to extract the mean
density of a star from a set of observed frequencies.}
% aims heading (mandatory)
{In this paper, we seek to establish a new method for estimating stellar mean
density, which combines the simplicity of a scaling law while providing the
accuracy of an inversion technique.}
% methods heading (mandatory)
{We provide a framework in which to construct and evaluate kernel-based linear
inversions which yield directly the mean density of a star.  We then describe
three different inversion techniques (SOLA and two scaling laws) and apply them
to the \object{sun}, several test cases and three stars, \object{$\alpha$ Cen B},
\object{HD 49933} and \object{HD 49385}, two of which are observed by
CoRoT.}
% results heading (mandatory)
{The SOLA (subtractive optimally localised averages) approach and the
scaling law based on the surface correcting technique described by
Kjeldsen et al. (2008, ApJ, 683, L175) yield comparable results which can reach an accuracy of
$0.5\,\%$ and are better than scaling the large frequency separation.  The
reason for this is that the averaging kernels from the two first methods are
comparable in quality and are better than what is obtained with the large
frequency separation. It is also shown that scaling the large frequency
separation is more sensitive to near-surface effects, but is much less affected
by an incorrect mode identification.  As a result, one can identify pulsation
modes by looking for an $\l$ and $n$ assignment which provides the best
agreement between the results from the large frequency separation and those from
one of the two other methods.  Non-linear effects are also discussed as is the
effects of mixed modes.  In particular, it is shown that mixed modes bring
little improvement to the mean density estimates as a result of their poorly
adapted kernels.}
% conclusions heading (optional)
{}

\keywords{stars: fundamental parameters --
          asteroseismology --
          stars: oscillations (including pulsations) --
          stars: interiors}
          %stars: individual: \object{$\alpha$ Cen B}, \object{HD 49933}, \object{HD 49385} --
          %\object{sun}: helioseismology}

\maketitle

\section{Introduction}

Determining accurate stellar mass is crucial to several domains in
astrophysics.  Indeed, mass plays a dominant role in the evolution and final
fate of stars.  It is also a key parameter when characterising exoplanetary
systems since the masses of exoplanets, essential for determining whether they
are rocky or gaseous in nature, are in general determined with respect to that
of the star.  However, in spite of its importance, stellar mass is usually
difficult to obtain accurately for single stars and is frequently model
dependent.

A method which has often been used for estimating stellar mass is to compare the
position of a star in an Hertzsprung-Russell (HR) diagram with
evolutionary tracks for stellar models of different masses.  The main
difficulties with this method is its model dependence, the large error bars, and
the fact that there are regions in the HR diagram in which more than one stellar
evolution track goes through each and every point. A promising alternative is to
use asteroseismology.  Indeed, acoustic modes are a sensitive indicator of the
mean density of a star.  Current space missions CoRoT \citep[\eg][]{Michel2008}
and Kepler \citep[\eg][]{Kjeldsen2010} are measuring stellar pulsation
frequencies to high accuracy.  When combined with an independent determination
of the radius, this yields the stellar mass.  Thanks to very precise parallax
measurements, radii accurate to $2\,\%$ are expected from the forthcoming
astrometric mission GAIA \citep{Perryman2001}. The main challenge is then
finding the best way of extracting the mean density of a star from a given set
of frequencies.

Various methods have been devised or proposed for estimating stellar mean
density from pulsation frequencies.  These include using simple scaling laws
based on the large frequency separation \citep{Ulrich1986} or other frequency
combinations \citep{Kjeldsen2008}, searching for the best fitting model within a
grid of models or a parameter space and retaining its mean density
\citep[\eg][]{Bazot2008, Metcalfe2010, Kallinger2010}, and performing a full
structural inversion to determine the density profile, then integrating it to
obtain the mean density.   The structural inversion may be carried out using
linear or non-linear techniques \citep[\eg][]{Gough1991, Antia1996,
Roxburgh2002}.  The scaling relations have the obvious advantage of simplicity
over the other methods and require fewer frequencies than a full structural
inversions, but are less accurate.  Although more expensive computationally,
searching for best fitting models in a grid is also straightforward and usually
more precise, but is model dependent, so that the conclusions are only as firm
as the input physics.  Full structural inversions have the advantage of being
less model dependent, but usually involve a number of free parameters which need
careful adjustment and can fail to converge when non-linear.  However, with the
current missions CoRoT and Kepler and future asteroseismic projects,
there is a growing need to provide accurate seismic stellar masses for a large
number of stars in an automatic way.  An ideal method would combine the
advantages of the above techniques while avoiding their weaknesses.

In this paper, we focus on kernel-based linear inversion methods which yield
directly the mean density without inverting for the full density profile.  As
such, this represents a first attempt at constructing methods which combine the
simplicity of scaling laws with the accuracy of full structural inversions. 
Although the results are not as accurate as initially hoped for, this approach
gives insight into the precise relation between the mean density of a star and
pulsation modes, and provides a better way of evaluating scaling laws. 
Section~\ref{sect:general_aspects} establishes a general framework in which to
construct and evaluate linear inversion techniques which yield the mean density
directly.  The following section then describes various linear inversion
techniques, including linearised scaling laws.  This is followed by a section
which shows how to extend these techniques beyond the linear regime. 
Sections~\ref{sect:Sun},~\ref{sect:grid} and~\ref{sect:observed_stars} applies
these different methods to the \object{sun}, to test cases using a grid of
models, and to three stars, two of which were observed by CoRoT  (\object{HD
49933} and \object{HD 49385}).  Finally a discussion concludes the paper.

\section{General aspects}
\label{sect:general_aspects}

\subsection{Starting point}
As for most inversion methods, the methods described below start with an
observed star and a reference stellar model.  At this point we will assume that
the star and the model have the same radius, but will relax this constraint
later on.  The structure of the reference model needs to be sufficiently close
to the structure of the star so that the link between the relative frequency
differences, $\delta\nu_{n\l}/\nu_{n\l}$, and the differences in the
stellar structure can be described by a linear relation:
\begin{eqnarray}
\frac{\delta\nu_{n\l}}{\nu_{n\l}} &=&
  \int_0^1 K_{\rho,\Gamma_1}^{n\l}(x) \frac{\delta \rho}{\rho} \d x
+ \int_0^1 K_{\Gamma_1,\rho}^{n\l}(x) \frac{\delta \Gamma_1}{\Gamma_1} \d x \nonumber \\
&+& \frac{F_{\mathrm{surf}}(\nu_{n\l})}{Q_{n\l}},
\label{eq:struct_problem}
\end{eqnarray}
where $\rho$ is the density, $\Gamma_1$ the first adiabatic exponent, $x=r/R$
the fractional radius, $n$ the radial order, $\l$ the harmonic degree, $R$ the
radius and $\frac{F_{\mathrm{surf}} (\nu_{n\l})} {Q_{n\l}}$ an ad-hoc surface
term \citep[\eg][]{Christensen-Dalsgaard1991}.  The function $F_{\mathrm{surf}}$
is assumed to be a slowly varying function of $\nu$ only, and $Q_{n\l}$
represents the mode's inertia \citep[\eg][]{Aerts2010}, normalised by the
inertia of a radial mode, interpolated to the same frequency.  The kernels
$K_{\rho,\Gamma_1}^{n\l}(r)$ and $K_{\Gamma_1,\rho}^{n\l}(r)$ are known
functions which can be deduced from the $(n,\l)$  eigenmode of the reference
model, via the variational principle \citep{Gough1991}.  Although the above
relation can also be applied to other structural pairs, we will work with the
pair $(\rho,\Gamma_1)$ since $\rho$ is an obvious choice for finding the mean
density, and $\Gamma_1$ is expected to vary little between different models. 
Throughout this paper, we will use the following definition for the relative
difference of a quantity $f$, although we note that other definitions have been
used \citep[\eg][]{Antia1994}:
\begin{equation}
\frac{\delta f}{f} = \frac{f_{\mathrm{obs}}-f_{\mathrm{ref}}}{f_{\mathrm{ref}}}.
\label{eq:delta_f}
\end{equation}
With this definition, we have the relation $f_{\mathrm{obs}} = f_{\mathrm{ref}}
(1 + \delta f/ f)$.

The mass difference, $\delta M$,  between the reference model and the star is
given by:
\begin{equation}
\delta M = \int_0^R 4\pi r^2 \delta \rho \d r.
\label{eq:mass}
\end{equation}
From this equation it is possible to deduce the relative difference in mean
density, $\bar{\rho} = 3M/4\pi R^3$:
\begin{equation}
\frac{\delta \bar{\rho}}{\bar{\rho}} = \int_0^1 4\pi x^2 \frac{\rho}{\rhoref}\frac{\delta \rho}{\rho} \d x,
\label{eq:mean_density}
\end{equation}
where the quantity $\rhoref = M/R^3$ is used for non-dimensionalisation.  It is
important to note that if $\delta \rho/\rho \equiv 1$ then $\delta
\bar{\rho}/\bar{\rho} = 1$, in other words:
\begin{equation}
1 = \int_0^1 4\pi x^2 \frac{\rho}{\rhoref} \d x.
\label{eq:integral_T}
\end{equation}

If we now assume that the radius of the reference model is different
from that of the star, then the model can be rescaled through homology,
\ie\ a transformation which preserves all dimensionless quantities, so
as to have the same radius as the star.  Furthermore, the mean density
needs to be kept the same, so as to avoid modifying the pulsation frequencies
when rescaling the model.  Equation~(\ref{eq:mass}) will then
give the relative mass difference between the star and the rescaled reference
model.  However, given their non-dimensional form and the fact that the density
is preserved between the original and scaled models,
Eqs.~(\ref{eq:struct_problem}) and~(\ref{eq:mean_density}) remain
unchanged and can be applied directly to the original (unscaled) reference
model, as long as $x=1$ corresponds to both the radius of the star, and the
radius of the reference model.  The procedures which will be described below
are based on these two equations, and therefore do not require
explicitly scaling the reference model to the same radius as the star.  They
will consequently yield the mean density rather than the mass, although the
later can be deduced if the radius is known.

\subsection{Linear inversions}

The procedures described in this paper then consist in searching for a linear
combination of the relative frequency differences that best reproduces the
relative mean density variation, $\delta \bar{\rho}/\bar{\rho}$:
\begin{equation}
\frac{\delta \bar{\rho}_{\mathrm{inv}}}{\bar{\rho}} = \sum_i c_i \frac{\delta\nu_i}{\nu_i},
\label{eq:inversion}
\end{equation}
where the index $i$ represents the pair $(n,\l)$.  Using
Eq.~(\ref{eq:struct_problem}) to replace $\delta\nu_i/\nu_i$ then yields the
following expression:
\begin{eqnarray}
\frac{\delta \bar{\rho}_{\mathrm{inv}}}{\bar{\rho}} &=& \int_0^1 \Kavg(x) \frac{\delta \rho}{\rho} \d x
                                    + \int_0^1 \Kcross(x) \frac{\delta \Gamma_1}{\Gamma_1} \d x \nonumber \\
                                   &+& \sum_i c_i \frac{F_{\mathrm{surf}}(\nu_i)}{Q_i},
\label{eq:mean_density_inv}
\end{eqnarray}
where:
\begin{eqnarray}
\Kavg(x)   &=& \mbox{averaging kernel}  = \sum_i c_i K_{\rho,\Gamma_1}^i(x),
\label{eq:Kavg} \\
\Kcross(x) &=& \mbox{cross-term kernel} = \sum_i c_i K_{\Gamma_1,\rho}^i(x).
\label{eq:Kcross}
\end{eqnarray}
In order for $\delta \bar{\rho}_{\mathrm{inv}}/\bar{\rho}$ to be a good estimate
of the relative mean density variation, the averaging kernel, $\Kavg$, needs to
be as close as possible to $4\pi x^2 \rho/\rhoref$, as can be seen by comparing
Eqs.~(\ref{eq:mean_density_inv}) and~(\ref{eq:mean_density}).  Furthermore,
$\Kcross$, which represents the cross-talk coming from relative variations on
the $\Gamma_1$ profile, needs to be reduced as does the last term, which
represents the surface effects.  Hence, these kernels provide a natural way of
evaluating an inversion procedure \citep[\eg][]{Christensen-Dalsgaard1990}.

\subsection{Various error bars}
\label{sect:error}

The starting point for obtaining the error on the mean density estimate is to
calculate the difference $(\delta \bar{\rho}_{\mathrm{inv}} - \delta
\bar{\rho}_{\mathrm{exact}})/\bar{\rho}$:
\begin{eqnarray}
\frac{\delta \bar{\rho}_{\mathrm{inv}} - \delta \bar{\rho}_{\mathrm{exact}}}{\bar{\rho}}
 &=& \int_0^1 \left(\Kavg - 4\pi \frac{\rho}{\rhoref} x^2 \right) \frac{\delta\rho}{\rho} \d x
  \nonumber \\
 &+& \int_0^1 \Kcross \frac{\delta \Gamma_1}{\Gamma_1} \d x
  + \sum_i c_i \varepsilon_i \nonumber \\
 &+& \sum_i c_i \frac{F_{\mathrm{surf}} (\nu_i)}{Q_i},
\label{eq:error}
\end{eqnarray}
where the $\varepsilon_i$ are the actual errors on the relative frequency
differences resulting from measurement errors.  If the true density and
$\Gamma_1$ profiles are known beforehand (for instance when testing out the
methods on stellar models rather than true stars) then it is possible to
quantify separately the different contributions to the error using
Eq.~(\ref{eq:error}).  Of course, in a realistic case, such profiles are not
accessible.

Using Eq.~(\ref{eq:error}), one can seek an upper bound on the error through the
Cauchy-Schwartz inequality:
\begin{eqnarray}
\left| \frac{\delta \bar{\rho}_{\mathrm{inv}} - \delta \bar{\rho}_{\mathrm{exact}}}{\bar{\rho}} \right|
&\leq& \left\| \Kavg - 4\pi \frac{\rho}{\rhoref} x^2 \right\|_2
       \left\| \frac{\delta \rho}{\rho}\right\|_2 \nonumber \\
   &+& \left\| \Kcross \right\|_2
       \left\| \frac{\delta \Gamma_1}{\Gamma_1} \right\|_2  \nonumber \\
   &+& \sqrt{\left(\sum_i c_i^2\right) \left(\sum_i \varepsilon_i^2\right)} \nonumber \\
   &+& \sqrt{\left(\sum_i c_i^2\right) \left(\sum_i \left(\frac{F_{\mathrm{surf}} (\nu_i)}{Q_i}\right)^2\right)},
\label{eq:error_upper_bound}
\end{eqnarray}
where $\|f\|_2 = \sqrt{\int_0^1 f^2(x) \d x}$.  The difficulty, of course, is
finding reasonable estimates for $\|\delta\rho/\rho\|_2$,
$\|\delta\Gamma_1/\Gamma_1\|_2$, and $\sqrt{\sum_i \left[F_{\mathrm{surf}}
(\nu_i)/Q_i\right]^2}$.  In their solar structural inversions,
\citet{Rabello-Soares1999} used the quantity $\|\Kavg - T\|_2$ as a measure of
how well the averaging kernels matched specified target functions, $T$ (which in
their case were localised around specific grid points), and $\|\Kcross\|_2$
to characterise how well cross-talk was being suppressed.

It is also useful to look at the statistical error on the estimated density
variation, resulting from frequency measurement error.  Let us denote as
$\sigma_i$ the $1\sigma$ error bar on the \textit{relative} frequency
differences, $\delta \nu_i/\nu_i$.  The resulting $1\sigma$ error bar on
the estimated mean density variation is then given by:
\begin{equation}
\sigma_{\delta \bar{\rho}/\bar{\rho}} = \sqrt{\sum_i c_i^2 \sigma_i^2},
\label{eq:one_sigma}
\end{equation}
where we have assumed that the errors on the individual frequencies are
independent.  The quantity $\sqrt{\sum_i c_i^2}$ is known as the \textit{error
magnification} and is the amount by which the error is amplified if the
$\sigma_i$ are uniform.

It is important to realise that there are further sources of error besides those
given in Eq.~(\ref{eq:error}).  These include non-linear effects where
Eq.~(\ref{eq:struct_problem}) is no longer valid, non-adiabatic effects which
are not included in the pulsation calculations nor in the variational principle
(used to calculate the kernels), inaccuracies in the structural kernels which
come from neglecting surface terms in the integration by parts (although
these may be absorbed up to some extent in $\sum_i c_i
F_{\mathrm{surf}}(\nu_i)/Q_i$) and numerical inaccuracies.

\subsection{Homologous transformations}
\label{sect:homologous}

Before going on to describe various inversion procedures, it is useful to point
out simple properties which can be deduced from homologous transformations. Let
us suppose that the observed star can be deduced from the model through a
homologous transformation, such that the relative density variation is
$\delta\bar{\rho}/\bar{\rho} = \epsilon$, where $\epsilon$ is a small
quantity. As is well known from dimensional analysis, the frequencies scale with
$\sqrt{GM/R^3}$.  Therefore, to first order the relative frequency variation
will be $\epsilon/2$.  As a consequence, a necessary and sufficient condition
to obtain the correct inversion result in the case of a homologous
transformation is:
\begin{equation}
2  = \sum_i c_i,
\label{eq:sum_c_i}
\end{equation}
where we have made use of Eq.~(\ref{eq:inversion}).  In what follows, inversion
procedures which satisfy this condition will be called ``unbiased''.

Furthermore, one can also deduce a simple property about the $K_{\rho,\Gamma_1}$
kernels.  Remembering that $\delta \Gamma_1/\Gamma_1 = 0$ in a homologous
transformation and making use of Eq.~(\ref{eq:struct_problem}) yields the
following result:
\begin{equation}
\frac{1}{2} \simeq \int_0^1 K_{\rho,\Gamma_1}^{i}(x)  \d x.
\label{eq:rhoGamma1Kernel}
\end{equation}
This relation is approximate because in deriving the kernel $K_{\rho,\Gamma_1}$,
there are several integrations by part in which surface terms are neglected
\citep{Gough1991}. Nonetheless, Eq.~(\ref{eq:rhoGamma1Kernel}) is satisfied to a
good degree of accuracy.

Using Eqs.~(\ref{eq:rhoGamma1Kernel}),~(\ref{eq:Kavg}) and~(\ref{eq:integral_T}), it is possible
to re-express Eq.~(\ref{eq:sum_c_i}) as follows:
\begin{equation}
\int_0^1 \Kavg (x) \d x \simeq \int_0^1 4 \pi x^2 \frac{\rho}{\rhoref} \d x,
\end{equation}
which is another way of expressing that an inversion procedure is unbiased.

\section{Inversion procedures}

We will now describe three different methods for estimating
the mean density of a star.  The first method is an inversion procedure which
yields directly inversion coefficients, $c_i$. The next two methods are
non-linear scaling laws rather than inversion procedures.  In order to put them
in the same form as an inversion procedure, we linearise these laws and obtain
associated coefficients $c_i$.  This then allows a systematic comparison of the three 
methods using the averaging and cross-term kernels.  Hence, although the second
and third method are not initially inversion procedures, they will be treated
and described as such in what follows.

\subsection{SOLA inversion}
The SOLA procedure \citep[subtractive optimally localised averages,][]
{Pijpers1992, Pijpers1994} consists in optimising $\Kavg$ by minimising it's
difference with a specified target function, $T$, and reducing the effects of
the remaining terms. This is achieved by minimising the following cost function:
\begin{eqnarray}
J(c_i) &=& \int_0^1 \left\{ \Kavg(x) - T(x) \right\}^2 \d x
        +  \beta \int_0^1 \left\{ \Kcross(x) \right\}^2 \d x \nonumber \\
       &+& \sum_{m=1}^{\Msurf} a_m \sum_i \frac{c_i \psi_m (\nu_i)}{Q_i}
        +  \lambda \left\{ 2 - \sum_i c_i\right\} \nonumber \\
       &+& \frac{\tan\theta \sum_i \left(c_i \sigma_i\right)^2}{\left<\sigma^2\right>},
\label{eq:SOLA}
\end{eqnarray}
where $\left< \sigma^2 \right> = \frac{1}{N} \sum_{i=1}^N \sigma_i^2$, $N$ being
the number of observed frequencies. In order to obtain the relative mean density
variation, the appropriate target function is:
\begin{equation}
T(x) = 4\pi x^2 \frac{\rho}{\rhoref}.
\label{eq:target}
\end{equation}
The first term on the right hand side of Eq.~(\ref{eq:SOLA}) is used to reduce
the differences between $\Kavg$ and $T$, and the next two terms minimise the
effects of cross-talk from $\Gamma_1$ and pollution from unknown near-surface
effects.  The functions $\psi_m$ are a basis of slowly varying
functions, such as the first $\Msurf$ Legendre polynomials, and the $a_m$ are
Lagrange multipliers which are used to suppress these terms.  The fourth term
ensures that the inversion is not biased, $\lambda$ being a Lagrange multiplier.
Finally, the last term is a regularisation term which reduces
$\sigma_{\delta\bar{\rho}/\bar{\rho}}$ (see Eq.~\ref{eq:one_sigma}).  The free
parameters in this approach are $\beta$ which regulates between optimising
$\Kavg$ and minimising $\Kcross$, $\theta$ which is used to adjust the amount of
regularisation, and $\Msurf$ which controls the number of terms used in
suppressing unknown surface effects.

Using Eq.~(\ref{eq:target}) to define the target function avoids inverting for
the entire density variation profile, $\delta \rho$, and integrating it to
obtain the mean density variation, which simplifies the calculations and reduces
the number of free parameters (such as the choice of a grid on which to carry
out the inversion, or the width of the target functions at each grid point). 
Furthermore, the averaging and cross-term kernels are calculated for the
inverted mean density variation, rather than for the density at specific grid
points, which is more useful in evaluating the inversion errors.  Finally, it is
worth noting that \citet{Pijpers1998} had already used a similar global approach
to determine the \object{sun}'s moment of inertia, and had found similar benefits.

\subsection{Large frequency separation}

An alternate approach is to use the well-known scaling law between the mean
density and the large frequency separation \citep[\eg][]{Ulrich1986}:
\begin{equation}
\left< \Delta \nu \right> \propto \sqrt{\bar{\rho}}.
\label{eq:scaling}
\end{equation}
where $\left< \Delta \nu \right>$ is an average of the large frequency
separation, $\nu_{n+1,\,\l} - \nu_{n,\,\l}$. This law has been applied by many
authors to obtain the mean density of a variety of stars
\citep[\eg][]{Kjeldsen1995, Mosser2010}. Recently, \citet{White2011} has
investigated its accuracy for a grid of models and shown that the error
is around $2$ to $3.5\,\%$ for stars with $\Teff$ below $6700$ K,
relevant to solar-like oscillations, but can reach $10\%$ in the worst cases,
\textit{when calibrated to solar values}.

Equation~(\ref{eq:scaling}) stems from the fact that the frequencies, and
consequently any linear combination thereof, also scale with $\sqrt{\bar{\rho}}$
for homologous transformations, as was pointed out in
Sect.~\ref{sect:homologous}.  This still approximately holds for non-homologous
transformations, although some dependence on the precise stellar structure will
appear.  Asymptotic analysis shows that for high order p-modes, $\left< \Delta
\nu \right> = \left[2\int_0^R \frac{dr}{c}\right]^{-1}$, \ie\ the inverse of the
time it takes for a sound wave to travel from the surface of a star to its
centre and back again \citep{Vandakurov1967}.  Given that sound waves travel
more slowly near the surface, the large frequency separation is sensitive to
surface conditions.

As such, this law is not directly comparable to the above procedure.  However,
one can apply this law in differential form, \ie\ chose a reference model which
is close to the star, calibrate the law using this model, and evaluate the mean
density of the star.  Given that the reference model is close to the star, one
can simply perturb Eq.~(\ref{eq:scaling}) to relate the relative variation of
the large frequency separation to the relative mean density variation:
\begin{equation}
2 \frac{\delta \left< \Delta \nu \right>}{\left< \Delta \nu \right>}
\simeq \frac{\delta \bar{\rho}}{\bar{\rho}}.
\label{eq:scaling_diff}
\end{equation}
If we assume that the average large frequency separation can be expressed
as a linear combination of the frequencies, $\left< \Delta \nu \right> = \sum_i
d_i \nu_i$, the left hand side of Eq.~(\ref{eq:scaling_diff}) becomes:
\begin{equation}
2 \frac{\delta \left< \Delta \nu \right>}{\left< \Delta \nu \right>}
= \frac{2 \sum_i d_i \delta \nu_i}{\sum_i d_i \nu_i}
= \frac{2 \sum_i d_i \nu_i \frac{\delta\nu_i}{\nu_i}}{\sum_i d_i \nu_i}
= \sum_i c_i \frac{\delta\nu_i}{\nu_i},
\end{equation}
where
\begin{equation}
c_i = \frac{2 d_i \nu_i}{\sum_j d_j \nu_j}.
\end{equation}
Equation~(\ref{eq:scaling_diff}) then takes on the form given in
Eq.~(\ref{eq:inversion}).  Consequently, it is also possible to construct
averaging and cross-term kernels for this approach, which can then be compared
with those coming from the SOLA approach.  By construction, this approach is
unbiased and will therefore provide the correct answer for homologous
transformations. One can also easily verify that $\sum_i c_i = 2$.

It important to realise that although the large frequency separation is
typically applied to p-modes in the asymptotic regime, Eqs.~(\ref{eq:scaling})
and~(\ref{eq:scaling_diff}) have a more general application.  Indeed, what is
important is the $\sqrt{\bar{\rho}}$ scaling which applies to all modes, not the
fact that the large separation asymptotically gives the inverse sound travel
time. Hence, unless specified otherwise, we will typically use all of the
available modes when applying Eqs.~(\ref{eq:scaling})
and~(\ref{eq:scaling_diff}).

In what follows, we will use the following estimate for the large frequency
separation in order to obtain the coefficients $d_i$:
\begin{equation}
\left< \Delta \nu \right> = \frac{\sum_{\l} \left(N_{\l}-1\right) \left< \Delta \nu(\l) \right>}
                                 {\sum_{\l} N_{\l}-1},
\label{eq:large_separation}
\end{equation}
where
\begin{eqnarray}
\left< \Delta \nu(\l) \right> &=& \frac{\sum_{i=1}^{N_{\l}} \left[\nu_i(\l) - \left< \nu(\l) \right>\right]
                                        \left[n_i(\l) - \left< n(\l) \right>\right]}
                                       {\sum_{i=1}^{N_{\l}} \left[n_i(\l) - \left< n(\l)\right>\right]^2}, \\
\left< \nu(\l) \right> &=& \frac{1}{N_{\l}} \sum_{i=1}^{N_{\l}} \nu_i(\l), \\
\left< n(\l) \right> &=& \frac{1}{N_{\l}} \sum_{i=1}^{N_{\l}} n_i(\l).
\end{eqnarray}
Here, $N_{\l}$ is the number of modes with a given harmonic degree, $\l$,
and the $n_i(\l)$ and $\nu_i(\l)$ are their associated radial orders and
frequencies, respectively. The above definition is a weighted average of the
least-square estimates of the mean large frequency separation from
\citet{Kjeldsen2008} for each $\l$ value.   Using a least-squares approach has
the advantage of not requiring consecutive radial orders to estimate the large
frequency separation.

\subsection{The KBCD approach}

Recently, \citet{Kjeldsen2008} proposed a method for correcting unknown surface
effects which they treated as a power law.  This method uses a
combination of the large frequency separation and an average of the frequencies
to estimate either the exponent of the power law or the mean density of the
star.  Having established the value of the exponent in the solar case,
\citet{Kjeldsen2008} then go on to use this value to estimate the mean density
of other stars.  In what follows, we will refer to this method as the
``KBCD approach''.  Like the scaling law based on the large frequency
separation, this approach is non-linear and therefore not directly comparable to
linear inversion methods.  However, if one assumes that the reference model is
close to the observed star, then it is possible to linearise this approach as
was done above.  To do so, we start with Eq.~(6) of \citet{Kjeldsen2008},
subtract 1 from both sides, and retain only first order terms:
\begin{equation}
\frac{1}{2} \frac{\delta\bar{\rho}}{\bar{\rho}} \simeq
\frac{b\frac{\delta\left<\nu\right>}{\left<\nu\right>} - \frac{\delta\left<\Delta\nu\right>}
{\left< \Delta\nu \right>}}{b-1}
\label{eq:Kjeldsen}
\end{equation}
where $b$ is the exponent involved in the power law describing the unknown
surface effects, and $\left<\nu\right>$ an average of the frequencies. 
Interesting limiting cases of the above equation are:
\begin{itemize}
\item[$\bullet$] $b=0$: this corresponds to fitting the large frequency
                 separation and brings us back to the previous section,
\item[$\bullet$] $b \rightarrow \infty$: this corresponds to fitting the average
                 frequency, $\left<\nu\right>$, directly.
\end{itemize}
One will also note that $b=1$ is a singular case which corresponds to a
degeneracy in the equations.  \citet{Kjeldsen2008} found $b=4.9$ for the
\object{sun} and consequently used this value for other stars.  We will do
likewise in what follows.

In practise, we will use Eq.~(\ref{eq:large_separation}) to obtain the large
frequency separation, and a similar equation to obtain the average frequency:
\begin{equation}
\left< \nu \right> = \frac{\sum_{\l} \left(N_{\l}-1\right) \left< \nu(\l) \right>}
                                 {\sum_{\l} N_{\l}-1},
\end{equation}
thereby extending this method to non-radial modes. Once more, it is
straightforward to see that this approach is unbiased.

\section{Non-linear extension}
\label{sect:pre_scaling}

The above procedures are linear and are therefore only adapted to reference
models which are sufficiently close to the observed star.  However, one may try
to extend their applicability by pre-scaling the reference model so as to
approach the mean density of the star and hopefully bring the relative
differences into the linear regime.  We therefore introduce a scale factor $s$,
such that the mean density of the scaled reference model is
$s^2\bar{\rho}_{\mathrm{ref}}$ and the corresponding frequencies
$s\nu^{\mathrm{ref}}_i$.  With this rescaled model, the frequency shifts then
take on the following expression:
\begin{equation}
\frac{\nu^{\mathrm{obs}}_i - s \nu^{\mathrm{ref}}_i}{s \nu^{\mathrm{ref}}_i}
= \frac{1}{s}\left(\frac{\delta\nu_i}{\nu_i} + 1\right) - 1.
\end{equation}
The inverted mean density becomes:
\begin{equation}
\bar{\rho}_{\mathrm{inv}}(s) = \bar{\rho}_{\mathrm{ref}} s^2 \left\{1 + \sum_i c_i 
\left[ \frac{1}{s}\left(\frac{\delta\nu_i}{\nu_i} + 1\right) - 1\right] \right\}.
\label{eq:rho_inv}
\end{equation}
If one assumes that the linear inversion procedure is unbiased, then Eq.~(\ref{eq:rho_inv})
can be simplified as follows, using Eq.~(\ref{eq:sum_c_i}):
\begin{equation}
\bar{\rho}_{\mathrm{inv}}(s) = \bar{\rho}_{\mathrm{ref}}
\left\{-s^2 + s\left(2 + \sum_i c_i \frac{\delta\nu_i}{\nu_i}\right)\right\}.
\label{eq:rho_inv_bis}
\end{equation}
Equation~(\ref{eq:rho_inv_bis}) is a 2$^{\mathrm{nd}}$ order equation with the
following maximum value:
\begin{equation}
\rhomax \equiv \bar{\rho}_{\mathrm{inv}} \left( \smax \right) = \bar{\rho}_{\mathrm{ref}} \smax^2,
\label{eq:rho_max}
\end{equation}
where
\begin{equation}
\smax = 1 + \frac{1}{2} \sum_i c_i \frac{\delta\nu_i}{\nu_i} = \frac{1}{2} \sum_i c_i 
\frac{\nu^{\mathrm{obs}}_i}{\nu^{\mathrm{ref}}_i}.
\label{eq:smax}
\end{equation}
It is this maximum value which we will use as the best mean density estimate.  
Indeed, if one were to start with a given reference model, obtain a first mean
density estimate through a given linear inversion, scale the reference model
using this mean density, re-apply the inversion to obtain a second mean
density,  re-scale the reference model, and iterate till convergence, then the
final mean density would be $\rhomax$.  Another way of looking at this is that
$\smax$ is the scale factor for which the inversion yields no further correction
to the mean density:
\begin{equation}
\sum_i c_i \left[ \frac{1}{\smax}\left(\frac{\delta\nu_i}{\nu_i} + 1\right) - 1\right]
= 0.
\end{equation}

It is then interesting to compare the result from a non-linear procedure to
the $\rhomax$ value obtained from the linearised version of the procedure. If we
start with the scaling law between the large frequency separation and the mean
density, then the $\rhomax$ value from the linearised procedure is:
\begin{equation}
\rhomax = \bar{\rho}_{\mathrm{ref}}
\left(1 + \frac{1}{2} \frac{2\delta\left<\Delta\nu\right>}{\left<\Delta\nu\right>}\right)^2
= \bar{\rho}_{\mathrm{ref}} \left(\frac{\left< \Delta\nu\right>_{\mathrm{obs}}}
{\left< \Delta\nu\right>_{\mathrm{ref}}}\right)^2,
\end{equation}
where we have used Eq.~(\ref{eq:delta_f}).  This is identical to what is given
directly by the original, non-linearised, scaling law, hence providing further
justification for choosing $\rhomax$ as the mean density estimate.  If we
calculate $\rhomax$ for the linearised version of the KBCD approach, the
result is:
\begin{equation}
\rhomax = \bar{\rho}_{\mathrm{ref}} \left(\frac{b\frac{\left<\nu\right>_{\mathrm{obs}}}
{\left<\nu\right>_{\mathrm{ref}}} - \frac{\left<\Delta\nu\right>_{\mathrm{obs}}}
{\left<\Delta\nu\right>_{\mathrm{ref}}}}{b-1}\right)^2.
\label{eq:rhomax_kjeldsen}
\end{equation}
This time the result is different from what is given by the original non-linear
procedure.  However, this difference turns out to be minute.  Indeed,
Eq.~(\ref{eq:rhomax_kjeldsen}) corresponds to the result which would be obtained
had the unknown surface effects been represented as a power law in terms of the
\textit{reference} frequencies rather than the \textit{observed} ones. Using
this formulation, the exponent obtained in the solar case becomes $4.86$ instead
of $4.89$, and mean density estimates are hardly affected (for $b$ fixed at
$4.9$). Hence, in what follows, we will use this formulation as representative 
of the KBCD approach.

Finally, the associated $1\sigma$ error bar on the estimated mean density is
now given by:
\begin{equation}
\sigma_{\rhomax} = \bar{\rho}_{\mathrm{ref}} \sigma_{\smax^2}
                 = \bar{\rho}_{\mathrm{ref}} \smax \sqrt{\sum_i c_i^2 \sigma_i^2},
\end{equation}
where we have assumed that $\sigma_i \ll 1$.  This is different
from what is given in Eq.~(\ref{eq:one_sigma}) due to the extra term $\smax$.

By construction, $\rhomax$ does not depend on the mean density of the initial
reference model as opposed to a linear inversion result (this can also be shown
explicitly through Eqs.~\ref{eq:rho_max} and~\ref{eq:smax}).  The same also
applies to $\sigma_{\rhomax}$.  Furthermore, the various individual kernels, the
averaging and cross-term kernels, and the target function remain unchanged since
they are non-dimensional and consequently do not depend on the scale factor.  As
such, they remain a useful diagnostic of the inversion procedure.  However, 
when calculating the error as based on Eq.~(\ref{eq:error}), the
relative difference on the density profile needs to take into account the scaling
by $\smax$:
\begin{equation}
\left(\frac{\delta\rho}{\rho}\right)_{\mathrm{scaled}} 
= \frac{1}{\smax^2} \left[
\left(\frac{\delta\rho}{\rho}\right)_{\mathrm{unscaled}} + 1 \right] - 1.
\end{equation}

\section{The solar case}
\label{sect:Sun}

\begin{figure}[htbp]
\includegraphics[width=\mycolumn]{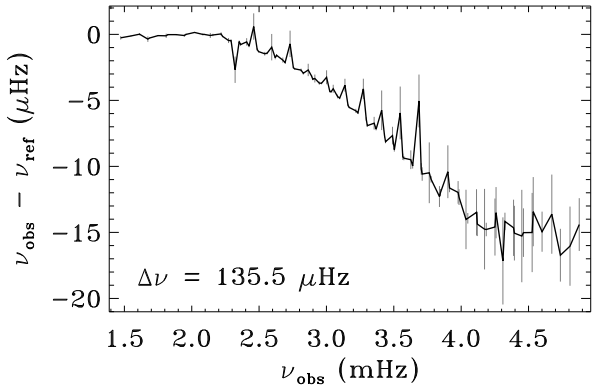}
\caption{Differences between observed solar cyclic frequencies and those obtained
from Model S, using the $\l=0-5$ frequency set from \citet{Lazrek1997}.  The vertical
grey lines indicate the $1\sigma$ observational error bars.  The large frequency
separation is also indicated.
\label{fig:sun_diff}}
\end{figure}

In order to illustrate the above methods, we apply them to the \object{sun}. A
set of $104$ GOLF frequencies and associated error bars \citep{Lazrek1997} were
used as observations. We only used the frequencies with a signal to noise ratio
greater than 1, \ie\ those which were not in brackets in Table~I of
\citet{Lazrek1997}. The $n$ and $\l$ values of these modes are given in
Table~\ref{tab:id_sun}, and the average of the error bars on these frequencies
is $0.694\,\mu$Hz.  Model S \citep{Christensen-Dalsgaard1996} was used as the
reference model. It is well known that there are surface effects in the
\object{sun} which are not accounted for in Model S.  This causes the high order
modes to differ in frequency from the observed values, as is illustrated in
Fig.~\ref{fig:sun_diff}.  For the sake of comparison, the large frequency
separation is also indicated on the figure. The pulsation modes were calculated
with ADIPLS \citep{Christensen-Dalsgaard2008}, using an isothermal outer
boundary condition, and the inversions were carried out using an improved (and
corrected) version of InversionKit\footnote{This software can be downloaded form
the Global Helioseismology HELAS website:\\
\href{http://helas.group.shef.ac.uk/science/inversions/InversionKit/index.php}
{\texttt{http://helas.group.shef.ac.uk/science/}}\\
\href{http://helas.group.shef.ac.uk/science/inversions/InversionKit/index.php}
{\texttt{inversions/InversionKit/index.php}}\\}.

\begin{table}[htbp]
\caption{The $n$ and $\l$ values of the 104 modes from \citet{Lazrek1997}.
\label{tab:id_sun}}
\begin{center}
\begin{tabular}{lcccccc}
\hline
\hline
$\l$ & 0 & 1 & 2 & 3 & 4 & 5 \\
$n$ & 10--34 & 9--34 & 10--31 & 13--30 & 14--24 & 18, 19 \\
\hline
\end{tabular}
\end{center}
\end{table}

Before giving the inversion results, it is useful to discuss what is the latest
and hopefully most accurate value of the solar mean density.  The mass and
radius used in Model S correspond to the values given in \citet{Allen1973}. 
Since then, the solar radius underwent a downward revision \citep{Brown1998}
following a discrepancy with the radius inferred from solar f-modes
\citep{Schou1997, Antia1998}.  Further revisions can be found in
\citet{Haberreiter2008} who also point out the need for a generally agreed upon
definition.  The solar mass is determined by calculating the ratio
$GM_{\odot}/G$, where $GM_{\odot}$ is known from planetary motions and $G$, the
gravitational constant, from various experiments.  The product $GM_{\odot}$ is
known to a high degree of accuracy so the main source of uncertainty comes from
the value of $G$.  A summary of the latest values as well as those used
in Model S can be found in Table~\ref{tab:sun}.  As can be seen in
Table~\ref{tab:sun} (fifth row), the latest mean density represents a
$0.14\pm0.09 \%$ increase over the value used in Model S.  However, given that
the pulsation frequencies scale with $\sqrt{GM/R^3}$, it is not possible to
distinguish between a variation on the solar mass and a modification of the
gravitational constant from frequency data alone.  Hence, it is more appropriate
to compare $G\bar{\rho}_{\odot}$ between Model S and the sun.  As can be seen in
the last row, the latest value represents a $0.16\pm0.08 \%$ increase over the
value from Model S.  This corresponds to what one would expect from inversions
if these were exact.  Given the excellent agreement between $GM_{\odot}$ from
Model S and the latest value (Table~\ref{tab:sun}, first row), the main
contribution to this difference comes from the discrepancy on the radius.

\begin{table}[htbp]
\caption{Solar mass and radius.
\label{tab:sun}}
\begin{center}
\begin{tabular}{lcc}
\hline
\hline
& Model S & Latest values \\
\hline
$GM_{\odot}\,\,\left(10^{26}\,\mathrm{cm}^{3}\mathrm{s}^{-2}\right)$ &
1.32712445\tablefootmark{a} &
1.32712440\tablefootmark{b} \\
$G\,\,\left(10^{-8}\,\mathrm{cm}^3\mathrm{g}^{-1}\mathrm{s}^{-2}\right)$ &
6.67232 &
6.67384(80)\tablefootmark{c} \\
$M_{\odot}\,\,\left(10^{33}\,\mathrm{g}\right)$&
$1.989$ &
$1.98855(24)$\tablefootmark{d} \\
%$1.98899993$\tablefootmark{a} \\
$R_{\odot}\,\,\left(10^{10}\,\mathrm{cm}\right)$ &
$6.9599$ & 
$6.95613(171)$\tablefootmark{e} \\
$\bar{\rho}_{\odot}\,\,\left(\gcm\right)$ &
1.4084 &
1.4104(12) \\
%1.4107(11) \\
$G\bar{\rho}_{\odot}\,\,\left(10^{-8}\,\mathrm{s}^{-2}\right)$ &
9.3975 &
9.4128(70) \\
\hline
\end{tabular} \\
\tablefoottext{a}{Based on the product $GM_{\odot}$ using the values
from the next two rows.}
\tablefoottext{b}{Based on \citet{Cox2000} and references therein.}
\tablefoottext{c}{2010 CODATA recommended value, taken from
\href{http://physics.nist.gov/cuu/Constants/index.html}{\texttt{http://physics.nist.gov/cuu/Constants/index.html}}}
\tablefoottext{d}{Based on the ratio $GM_{\odot}/G$ using the values
from the two previous rows.}
\tablefoottext{e}{Based on an average of the two values and quoted error bar in Table~3 of
\citet{Haberreiter2008}.}
\end{center}
\end{table}

\begin{figure*}[htbp]
\begin{tabular}{cc}
\includegraphics[width=\mycolumn]{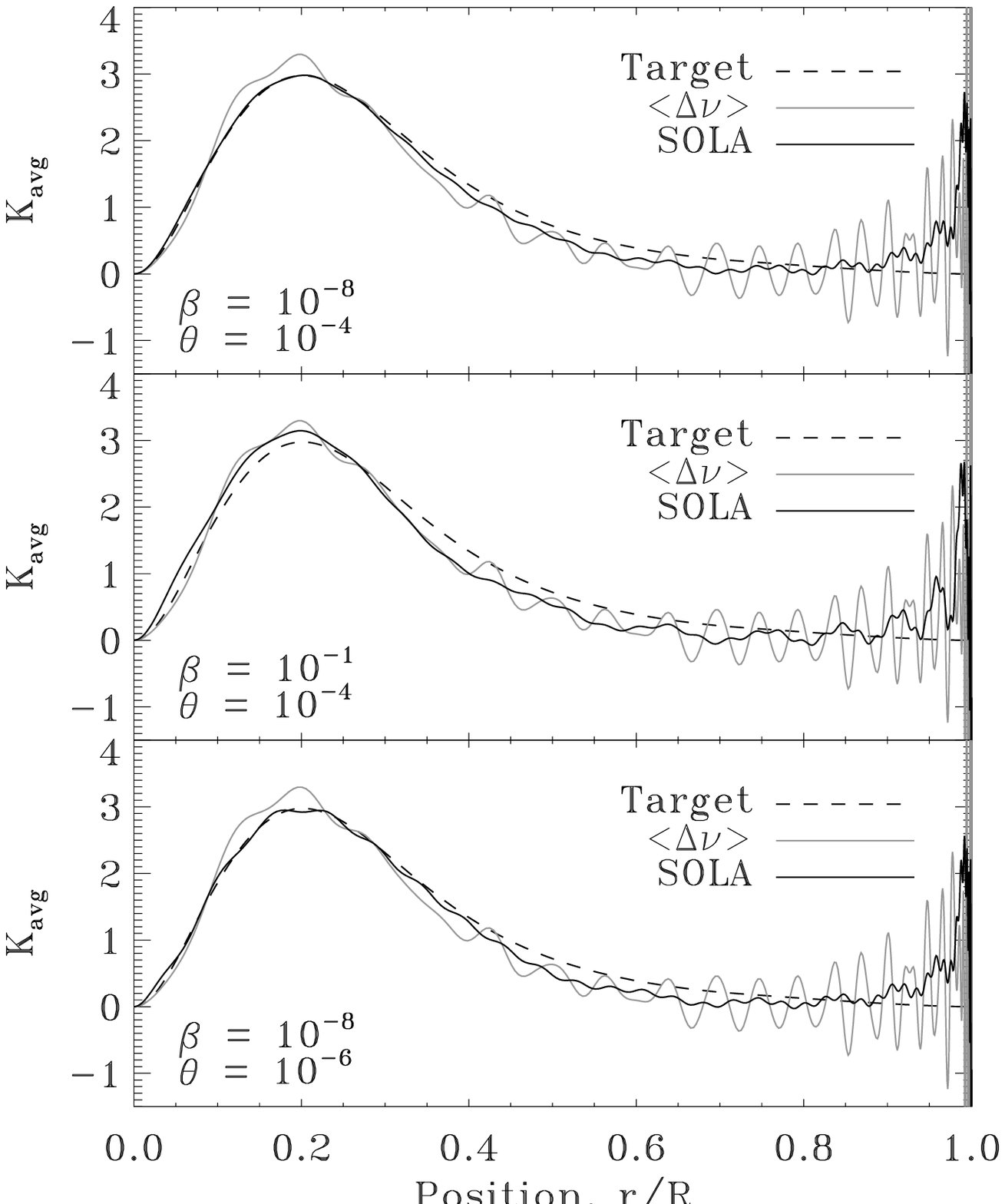} &
\includegraphics[width=\mycolumn]{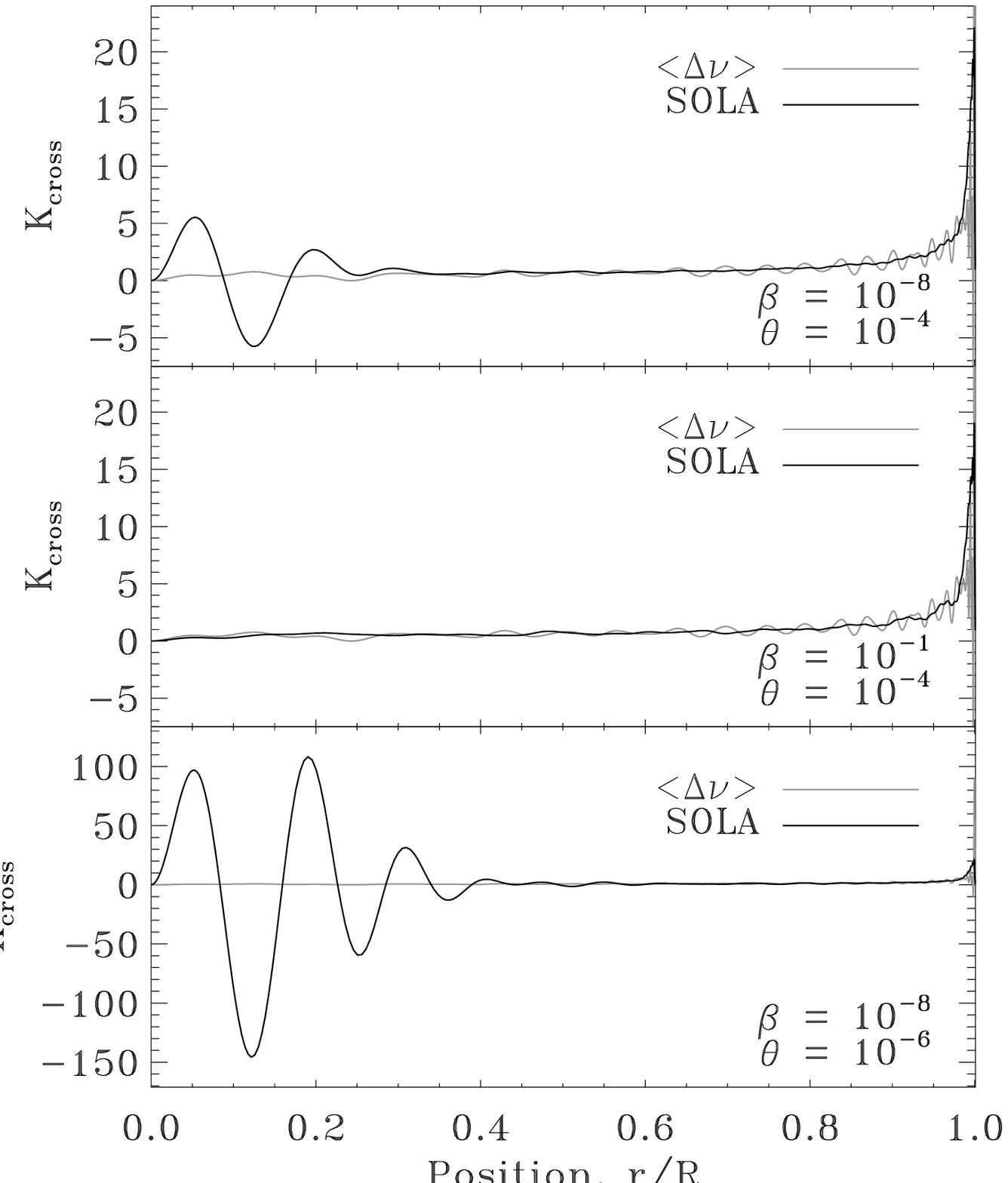}
\end{tabular}
\caption{Averaging and cross-term kernels from solar inversions for various
parameter settings.  No surface corrections are included in the inversion
($\Msurf = 0$).
\label{fig:sun_kernels}}
\end{figure*}

Table~\ref{tab:results_sun} lists the results for SOLA inversions with various
parameter settings, and for a $\left<\Delta\nu\right>$ scaling.  The
corresponding averaging and cross-term kernels are shown in
Fig.~\ref{fig:sun_kernels}.  The first two columns give the values of $\beta$
and $\theta$, when relevant.  No surface corrections are included in the SOLA
approach (\ie, $\Msurf=0$ in Eq.~\ref{eq:SOLA}).  The third column gives the
relative mean density variation (taking into account the non-linear
generalisation described in Sect.~\ref{sect:pre_scaling}).  As can be seen, all
of the inversions predict a slight decrease of the mean density, as opposed to
the $0.16\%$ increase expected from the latest solar parameters.  This is likely
to be caused by poorly modelled surface effects.  However, we do note that the 
$\left<\Delta\nu\right>$ scaling is much more affected, which is not surprising
given that the sound travel time integral is dominated by the near-surface
contribution.  The next column gives the $1\sigma$ error bar around
$\delta\bar{\rho}/\bar{\rho}$ resulting from the measurement uncertainties on
$\delta\nu_i/\nu_i$.  Finally the last two columns give $\|\Kavg-T\|_2$  and
$\|\Kcross\|_2$, which intervene in the upper bound on the error, as given in
Eq.~(\ref{eq:error_upper_bound}).

\begin{table*}[htbp]
\caption{Inversion results for the \object{sun}.
\label{tab:results_sun}}
\begin{center}
\begin{tabular}{cccccc}
\hline
\hline
$\beta$ &
$\theta$ &
$\delta\bar{\rho}/\bar{\rho}$ &
$\sigma_{\delta\bar{\rho}/\bar{\rho}}$ &
$\|\Delta \Kavg\|_2$ &
$\|\Kcross\|_2$ \\
\hline
$10^{-8}$ & $10^{-4}$ & $-1.9\times 10^{-3}$ & $5.3 \times 10^{-4}$ & 0.32 & 2.54 \\
$10^{-1}$ & $10^{-4}$ & $-5.1\times 10^{-3}$ & $1.9 \times 10^{-3}$ & 0.35 & 1.80 \\
$10^{-8}$ & $10^{-6}$ & $-1.2\times 10^{-3}$ & $1.2 \times 10^{-2}$ & 0.31 & 40.6 \\
\multicolumn{2}{c}{$\left<\Delta \nu\right>$ scaling} & $-1.2\times 10^{-2}$ & $4.1 \times 10^{-4}$ & 1.36 & 2.77 \\
\hline
\end{tabular}
\end{center}
\tablefoot{The first three rows correspond to SOLA inversions with different
parameter settings and no surface corrections ($\Msurf=0$), whereas the last row
corresponds to a $\left<\Delta\nu\right>$ scaling.  The associated kernels are
shown in Fig.~\ref{fig:sun_kernels}.}
\end{table*}

As for most inversion methods, the results are a trade-off between minimising the
effects of measurement errors and cross-talk from $\delta\Gamma_1/\Gamma_1$ and
optimising the averaging kernel.  The parameters $\beta = 10^{-8}$,
$\theta=10^{-4}$ seem to be a good compromise.  If $\beta$ is increased to
$10^{-1}$ then the inversion wastes much effort trying to reduce the cross-talk
from $\delta\Gamma_1/\Gamma_1$, which is expected to be small.  Although
$\|\Kcross\|_2$ is somewhat reduced, the inversion result is worse and
$\sigma_{\delta\bar{\rho}/\bar{\rho}}$ has noticeably increased.  If instead,
$\theta$ is reduced to $10^{-6}$, the inversion result is slightly better as is
$\|\Kavg-T\|_2$, but both $\sigma_{\delta\bar{\rho}/\bar{\rho}}$ and
$\|\Kcross\|_2$ take on large values.

\subsection{Near-surface effects}

\begin{figure*}[htbp]
\begin{tabular}{cc}
\includegraphics[width=\mycolumn]{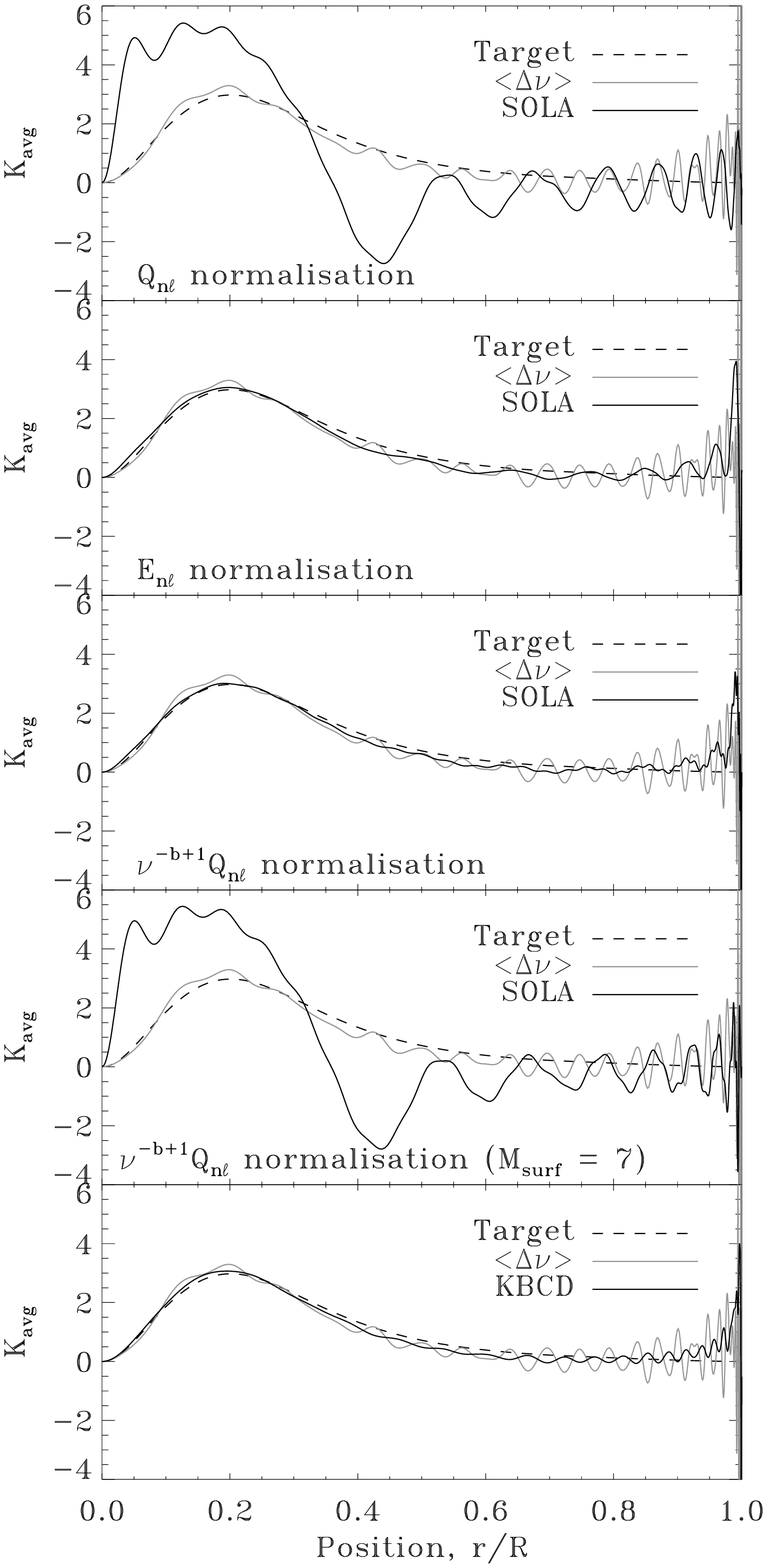} &
\includegraphics[width=\mycolumn]{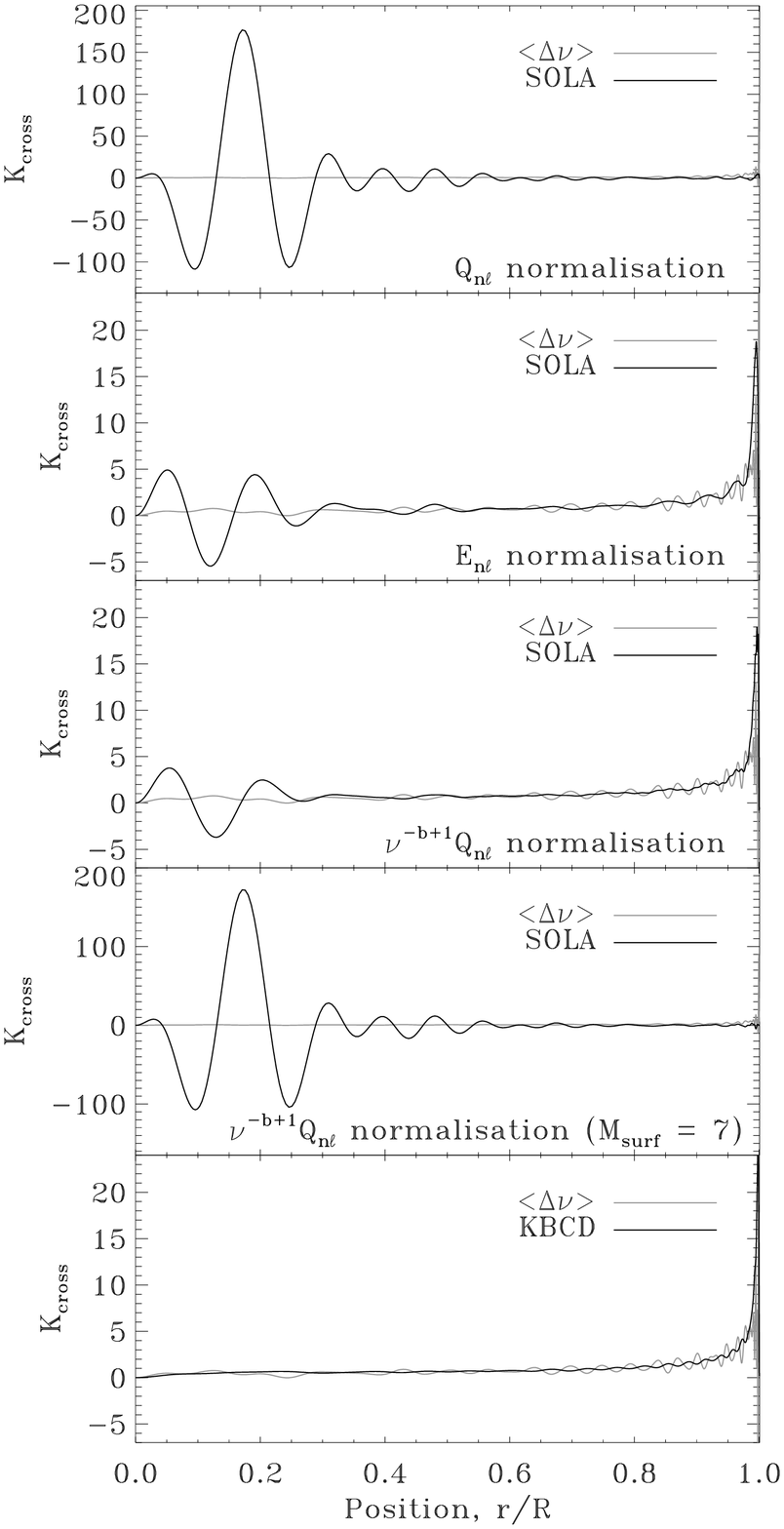}
\end{tabular}
\caption{Averaging and cross-term kernels from solar inversions for different
treatments of surface effects.  The last row shows the averaging and
cross-term kernels from the KBCD approach rather than from a SOLA inversion.
\label{fig:sun_kernels_surface}}
\end{figure*}

It is then interesting to look at different ways of correcting surface effects.
Traditionally, these effects have been modelled as an additive term made up of a
unknown function of frequency only, $F_{\mathrm{surf}}$, normalised by $Q_i$,
the mode inertia divided by the inertia of a radial mode interpolated to the
same frequency \citep{Christensen-Dalsgaard1991}.  One way of dealing with this
term in a SOLA inversion is to treat the function $F_{\mathrm{surf}}$ as a
slowly varying function of $\nu$, such as a linear combination of the first
$\Msurf$ Legendre polynomials, and then to constrain the inversion to cancel out
\textit{any} such function \citep[][also see Eq.~(\ref{eq:SOLA})]{Dappen1991}. 
The first row of Table~\ref{tab:results_sun_surface} gives the inversion result
and characteristics for $\Msurf = 1$, which amounts to treating
$F_{\mathrm{surf}}$ as a constant. The corresponding averaging and cross-term
kernels are given in the top panels of Fig.~\ref{fig:sun_kernels_surface}.  As
can be seen, the results are quite poor.  A likely reason for this behaviour is
that the function $F_{\mathrm{surf}}$ is not well described by a constant.
Instead, judging from Fig.~\ref{fig:sun_diff}, and remembering that $Q_i$ is
normalised in such a way as to be close to 1 for low $\l$ values,
$F_{\mathrm{surf}}$ starts at small values and increases with frequency. 

\begin{table*}[htbp]
\caption{Solar inversion results using different schemes for correcting surface
effects.
\label{tab:results_sun_surface}}
\begin{center}
\begin{tabular}{cccccc}
\hline
\hline
Description &
$\Msurf$ &
$\delta\bar{\rho}/\bar{\rho}$ &
$\sigma_{\delta\bar{\rho}/\bar{\rho}}$ &
$\|\Delta \Kavg\|_2$ &
$\|\Kcross\|_2$ \\
\hline
%     $10^{-8}$ & $10^{-4}$ & $-1.9\times 10^{-3}$ & $5.3 \times 10^{-4}$ & 0.32 & 2.54 \\ no surface corrections
$Q_{n\l}$ normalisation & 1 & $-5.6\times 10^{-2}$ & $2.2 \times 10^{-2}$ & 1.84 & 47.0 \\
$E_{n\l}$ normalisation & 1 & $-1.0\times 10^{-3}$ & $1.2 \times 10^{-3}$ & 0.50 & 2.55 \\
$\nu^{-b+1}Q_{n\l}$ normalisation & 1 & $-8.5\times 10^{-4}$ & $6.2 \times 10^{-4}$ & 0.40 & 2.25 \\
$\nu^{-b+1}Q_{n\l}$ normalisation & 7 & $-5.8\times 10^{-2}$ & $2.3 \times 10^{-2}$ & 1.86 & 45.9 \\
\citet{Kjeldsen2008}    & -- & $-1.9\times 10^{-3}$ & $5.5 \times 10^{-5}$ & 0.41 & 2.03 \\
%$\nu^{-b}Q_{n\l}$ normalisation & 1 & $-9.6\times 10^{-4}$ & $5.4 \times 10^{-4}$ & 0.38 & 2.27 \\
%$\nu^{-b}Q_{n\l}$ normalisation & 6 & $-4.3\times 10^{-2}$ & $1.7 \times 10^{-2}$ & 1.63 & 29.3 \\
%$\nu^{-b+1}Q_{n\l}$ normalisation & 6 & $-5.6\times 10^{-2}$ & $2.2 \times 10^{-2}$ & 1.82 & 41.9 \\
%Frequency ratios        & -- & $1.34$               & $4.6 \times 10^{-1}$ & 1.14 & 411  \\
\hline
\end{tabular}
\end{center}
\tablefoot{The parameters $\beta = 10^{-8}$, $\theta=10^{-4}$ were used for the
SOLA inversions (first four rows).  The associated kernels are shown in
Fig.~\ref{fig:sun_kernels_surface}.}
\end{table*}

One way of trying to deal with this problem is to normalise the function
$F_{\mathrm{surf}}$ in a different way, before treating it as a linear
combination of slowly varying functions.  Two alternatives are:
\begin{equation}
\frac{G_{\mathrm{surf}} (\nu_i)}{E_i}, \qquad \mbox{and} \qquad
\frac{H_{\mathrm{surf}} (\nu_i)}{\nu_i^{-b+1} Q_i},
\end{equation}
where we have used the notation $G_{\mathrm{surf}}$ and $H_{\mathrm{surf}}$ to
distinguish these functions from the original function $F_{\mathrm{surf}}$, and
where $E_i$ is the mode inertia and $b$ the exponent from the KBCD approach. 
The first possibility is based on the following reasoning.  In the original
description, the quantity $Q_i$ was introduced to replace $E_i$ as it doesn't
vary by several orders magnitude over the p-mode frequency range
\citep{Christensen-Dalsgaard1991}.  However, factoring out this strong variation
from $E_i$ also means introducing it into $F_{\mathrm{surf}}$.  Therefore,
normalising by $E_i$ instead of $Q_i$ avoids this problem.  The second
possibility is based on the idea that $F_{\mathrm{surf}}(\nu_i) = Q_i
(\delta\nu_i/\nu_i)_{\mathrm{surf}} \simeq a\nu_i^{b-1}$, based on
\citet{Kjeldsen2008}.  Hence by dividing both sides by $\nu^{b-1}$, one is left
with a function $H_{\mathrm{surf}}$ which is slowly varying.

Rows 2-4 of Table~\ref{tab:results_sun_surface} give the results from applying
these two alternative normalisations, and the corresponding kernels are shown in
Fig.~\ref{fig:sun_kernels_surface}.  In the first two cases, only one (constant)
term is used to correct surface effects.  The results are slightly better than
when no surface corrections are included, with a slight preference for the
$\nu^{-b+1} Q_i$ normalisation.  At the same time, both the averaging kernel and
$\sigma_{\delta\bar{\rho}/\bar{\rho}}$ are slightly worse.  In last case,
$\Msurf=7$ in accordance with the optimal number of terms found in
\citet{Rabello-Soares1999}. However, this leads to poor results.  The main
difference between the inversions carried out here and those carried out in
\citet{Rabello-Soares1999} is the number of modes involved in the inversion.
Given the limited number used here (which still remains generous in comparison
with stars other than the \object{sun}), one cannot afford to over-correct for surface
effects.  However, even when no surface corrections are explicitly included,
Table~\ref{tab:results_sun} shows that a SOLA inversion is less affected by
surface effects than a $\left<\Delta\nu\right>$ scaling, probably because as
$\Kavg$ approaches the target function $T$, its surface amplitude is reduced.

Another way to try to limit surface effects is to perform a SOLA inversion using
only kernels based on frequency separation ratios.  Indeed, \citet{Roxburgh2003}
showed that these ratios are insensitive to surface conditions through
analytical developments based on internal phase shifts and through numerical
simulations.  Likewise, \citep{OtiFloranes2005} showed that the associated
kernels have a small amplitude near the surface.  However, they also pointed out
that both the numerator and denominator of a frequency separation ratio scale
with $\sqrt{GM/R^3}$, meaning that the ratio is insensitive to the mean density of
the underlying model.  As a result, the $c_i$ coefficients associated with an
inversion based on these kernels add up to $0$, making it impossible to do an
unbiased inversion.  Likewise, it can be shown using
Eq.~(\ref{eq:rhoGamma1Kernel}) that the integrals of the kernels
$K_{\rho,\Gamma_1}^{\mathrm{ratio}}$  is approximately $0$.

%We note, in passing, that alternate ways of dealing with surface effects have
%been proposed \citep[\eg][]{Basu1996}.

Finally, the last row of Table~\ref{tab:results_sun_surface} gives the results
from applying the KBCD approach.  The associated averaging and
cross-term kernels are displayed in the lower panels of
Fig.~\ref{fig:sun_kernels_surface}.  The mean density estimate and
$\sigma_{\delta\bar{\rho}/\bar{\rho}}$ compare favourably with the other
inversion results and the associated kernels are well behaved, thus explaining
why this method works.  This approach is found to be superior to the simple
scaling law based on the large frequency separation, as can be seen both in
terms of the results it produces and in terms of the kernels.  The likely reason
for this is that the averaged frequency, which intervenes in this approach, uses
a more uniform weighting on the individual frequencies whereas the large
frequency separation favours frequencies at either end of the frequency range,
thus amplifying surface effects which are strongest at high radial orders.
Overall, the best results are obtained with the $\nu^{-b+1}Q_i$
normalisation, using one term for surface effects, and the KBCD approach
(3$^\mathrm{rd}$ and 5$^\mathrm{th}$ line of
Table~\ref{tab:results_sun_surface}).

\section{Test cases with a grid of models}
\label{sect:grid}

\begin{figure}[htbp]
\begin{center}
\includegraphics[width=\mycolumn]{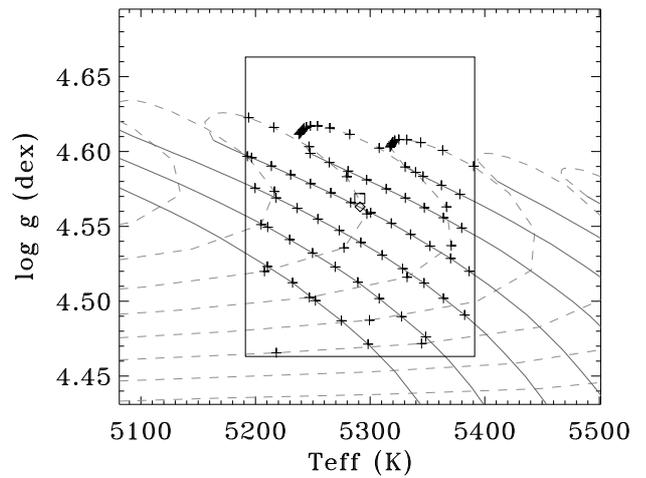}
\caption{$\Teff$ - $\logg$ diagram showing the position of Model A ($\Diamond$),
Model B ($\square$), the reference models from grid B ($+$), and various
pre-main-sequence and main-sequence tracks, represented by dashed and continuous
line respectively.
\label{fig:Teff_logg1}}
\end{center}
\end{figure}

\begin{figure}[htbp]
\includegraphics[width=\mycolumn]{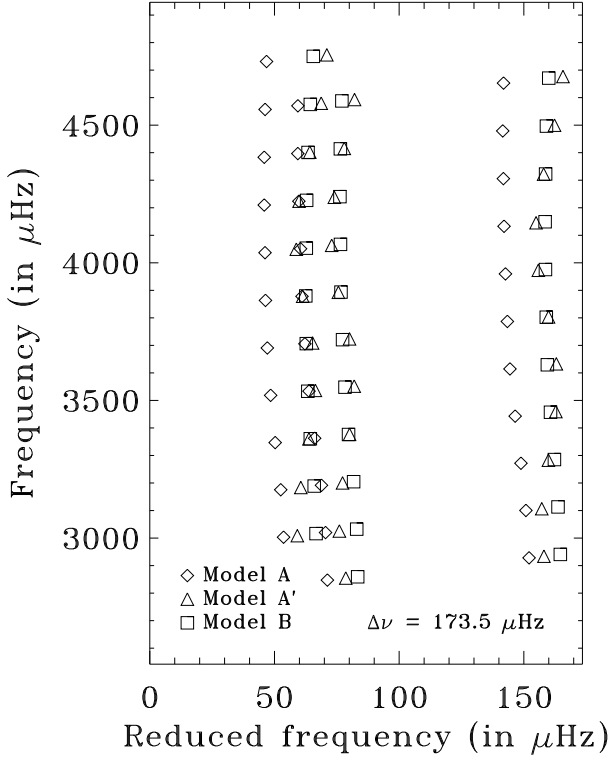}
\caption{Echelle diagram with the frequencies from Models A, \Aprime\ and B. The
$\l=2$ and $\l=0$ ridges are to the left, whereas the $\l=1$ ridge is to the
right. \textit{Note:} the large frequency separation indicated in the lower
right corner is the value used to construct this figure, not necessarily the
exact value for the three models.
\label{fig:echelle1}}
\end{figure}

In order to do a more systematic study, we apply the different inversion
procedures to a grid of 93 main-sequence and pre-main-sequence models.  These
models were downloaded from grid B of the CoRoT-ESTA-HELAS
website\footnote{\URL{http://www.astro.up.pt/corot/models/cesam/}}, and were
chosen within the $(T_{\mathrm{eff}},\log(g))$ box $(5291 \pm
100\,\mathrm{K},4.563 \pm 0.1\,\mathrm{dex})$, the error bars being
representative of what can be expected from spectroscopic observations.  
Their masses range from $0.80\,M_{\odot}$ to $0.92\,M_{\odot}$ and their
ages from $28$ Myrs to $17.6$ Gyrs, after their appearance on the birthline.  A
description of the physical ingredients, chemical composition and initial
conditions used in these models is given in \citet{Marques2008}.  In particular,
the initial chemical composition is $(X,Y,Z) = (0.70,0.28,0.02)$, where the
metal abundances correspond to the solar mixture given in \citet{Grevesse1993}. 
Figure~\ref{fig:Teff_logg1} shows the positions of these models in a
$(T_{\mathrm{eff}},\log(g))$ diagram, along with some evolutionary tracks, the
dashed parts corresponding to the pre-main-sequence.  Three additional models,
nicknamed ``Models A, \Aprime\ and B'', were selected near the centre of this
box to play the role of observed stars.  These models are also shown in
Fig.~\ref{fig:Teff_logg1}, and Table~\ref{tab:ModelsAB} gives some of the
characteristics of Models A and B.  The frequencies of these models are shown in
an {\'e}chelle diagram in Fig.~\ref{fig:echelle1}.  In all the cases tested here,
the $1\sigma$ error bar on the frequencies was set at $0.3,\,\mu$Hz, although 
no errors were added to the frequencies.  The goal of this study is then to see
how well the density of these three models can be reproduced by applying the
various inversion procedures to the models from the grid.

\begin{table}[htbp]
\begin{center}
\caption{Characteristics of Models A and B.
\label{tab:ModelsAB}}
\begin{tabular}{lcc}
\hline
\hline
& Model A & Model B \\
\hline
$\bar{\rho}$ ($\gcm$) & 2.2895  & 2.3079 \\
Mass ($M_{\odot}$)    & 0.900   & 0.920  \\
Radius  ($R_{\odot}$) & 0.821   & 0.825  \\
Age (Gyr)             & 1.492   & 2.231  \\
$\logg$ (dex)         & 4.563   & 4.569  \\
$\Teff$ (K)           & 5291    & 5291   \\
%$\Teff$ (K)          & 5291.2  & 5290.9 \\
\hline
\end{tabular}
\tablefoot{Here, $M_{\odot}$ and $R_{\odot}$ take on the values from Model S.}
\end{center}
\end{table}

\subsection{Model A}

Model A represents a $0.9\,M_{\odot}$ main-sequence star and was downloaded from
grid A of the CoRoT-ESTA-HELAS website\footnote{The specific URL to Model A is\\
\href{http://www.astro.up.pt/corot/models/cesam/A/data/0.90/ms/star\_1492.002-ad.osc.gz}
{\texttt{http://www.astro.up.pt/corot/models/cesam/}}\\
\href{http://www.astro.up.pt/corot/models/cesam/A/data/0.90/ms/star\_1492.002-ad.osc.gz}
{\texttt{A/data/0.90/ms/star\_1492.002-ad.osc.gz}}}.
Models from grid A are started off as fully convective spheres, instead of using
a birthline as in grid B.  Apart from this difference, this model has the same
chemical composition and physical ingredients as the models in the grid.

\begin{figure*}[htbp]
\includegraphics[width=\textwidth]{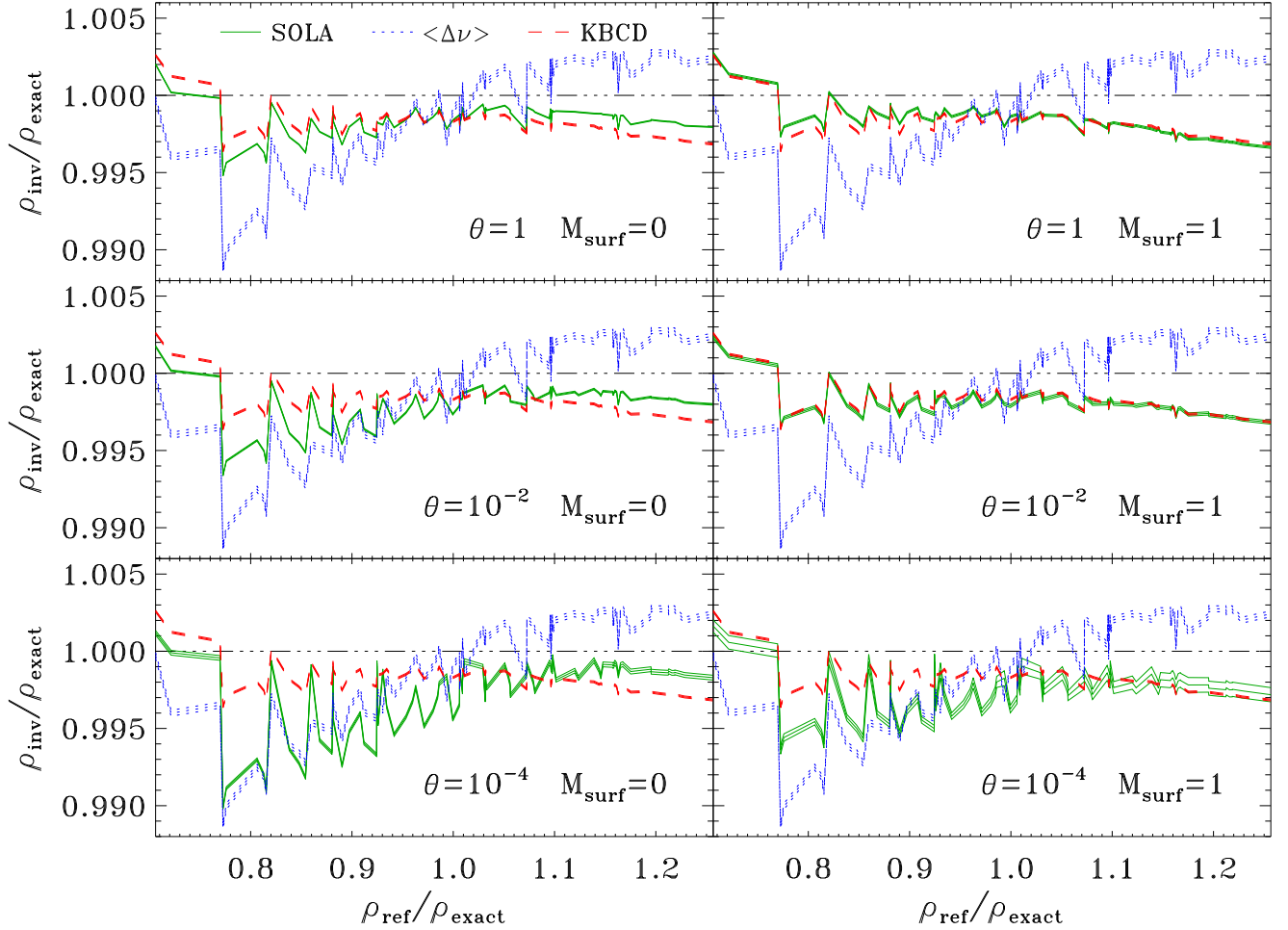} 
\caption{Inversion results for Model A, using the modes with $n=15-25$ and
$\l=0-2$.  The $x$ axis gives the mean density of the reference models
divided by that of Model A, whereas the $y$ axis is the mean densities obtained
from the various inversion procedures, normalised by that of Model A.  For each
procedure, there are three lines corresponding to the result and the $1\sigma$
error bar around it. The horizontal triple-dot-dashed line shows the correct
mean density.  A $\nu^{-b+1}Q_{n\l}$ normalisation has been used when surface
corrections are included in SOLA inversions.
\label{fig:ModelA}}
\end{figure*}

Each one of the models from the grid was used as a reference model from
which to apply the different inversion techniques to the ``observed''
frequencies from Model A.  Figure~\ref{fig:ModelA} shows the results from
these inversions for a set of $33$ modes with radial orders $n=15-25$
and harmonic degrees $\l=0-2$.  The parameter $\beta$ was set to
$10^{-8}$ for all of the SOLA inversions.  Although the mean densities of the
reference models differ from that of Model A by up to $30\,\%$ (see the $x$-axis
in figures), the inversion results are within $0.3$ to $0.6\,\%$ of the
true mean density except for the worst cases where the departure can
exceed $1.0\,\%$, which shows the sensitivity of these modes to density.  A
more detailed look at the figures shows that overall, the KBCD approach
and SOLA inversions produce the best results, provided the parameter
$\theta$ is not too small.  The two are very similar if $\Msurf=1$,
$\theta=10^{-2}$ and a $\nu^{-b+1}Q_{n\l}$ normalisation is used for surface
effects.  The large frequency separation, for the most part, produces results
which are worse.  One may also notice that in most cases, the difference
between the inverted and true mean densities is much larger than the $1\sigma$
error bar from the inversions.  This behaviour can be expected, since the 
$1\sigma$ error bar from the inversion is only based on the frequency measurement
errors and does not take into account the other sources of error mentioned in
Sect.~\ref{sect:error}, such as differences between the averaging kernel and the target
function or cross-talk from the $\delta\Gamma_1/\Gamma_1$ profile.

\begin{figure}[htbp]
\includegraphics[width=\mycolumn]{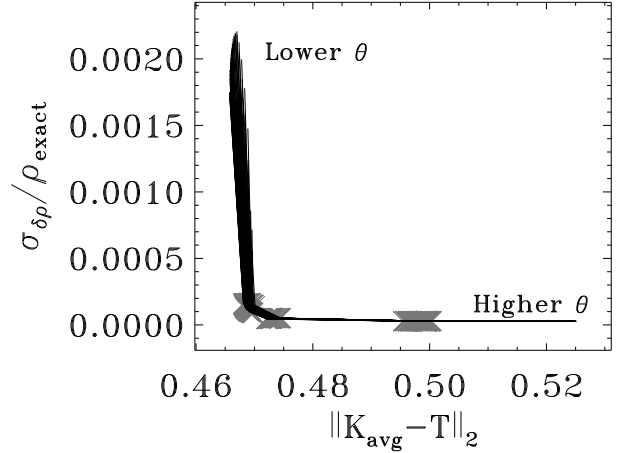} 
\caption{Superimposed trade-off curves for each of the 93 models from the grid. 
The parameters are $\Msurf=0$, $\beta=10^{-8}$ and $\theta=10^{-6},\,10^{-4},\,
10^{-2},\,1,\,1.5$. The grey ``$\times$'' marks the locations of intermediate
$\theta$ values.
\label{fig:lcurve}}
\end{figure}

As can be seen, reducing $\theta$ mainly seems to increase the $1\sigma$ error
bar in SOLA inversions without improving the results.  This can be understood by
looking at the trade-off curves illustrated in Fig.~\ref{fig:lcurve}.  Indeed,
reducing $\theta$ only has a small effect on $\|\Kavg-T\|_2$ while
$\sigma_{\delta\bar{\rho}/\bar{\rho}}$ increases substantially.  Furthermore, 
the same mechanism which amplifies observational errors can also
amplify errors which are not accounted for in Eq.~(\ref{eq:error}), such
as non-linear effects not included in Eq.~(\ref{eq:struct_problem}) or numerical
errors on kernel calculations.  Based on Figs.~\ref{fig:ModelA}
and~\ref{fig:lcurve}, the value $\theta=10^{-2}$ seems to be a good compromise
between the two effects.

\begin{figure}[htbp]
\includegraphics[width=\mycolumn]{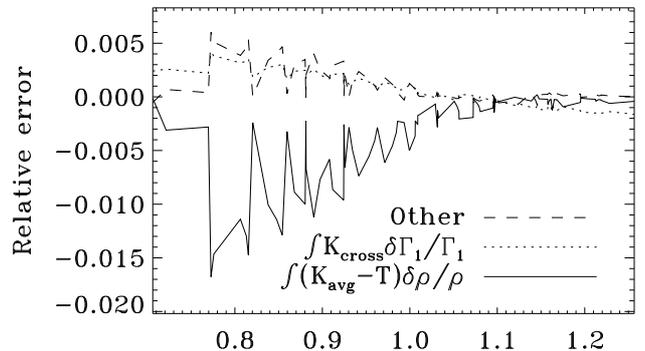} 
\caption{Different sources of error based on Eq.~(\ref{eq:error}) for SOLA
inversions.  See text for details (in particular for ``Other''). The SOLA
parameters are $\beta=10^{-8}$, $\theta=10^{-2}$ and $\Msurf=0$.
\label{fig:error}}
\end{figure}

Furthermore, given that the true structure of Model A is known, it is
possible to analyse the different sources of error based on
Eq.~(\ref{eq:error}).  These are displayed in Fig.~\ref{fig:error} for SOLA
inversions.  The dominant contribution comes from a mismatch between the
averaging kernel and the target function and is represented by a solid line. 
The dotted line represents the cross-talk from the $\delta\Gamma_1/\Gamma_1$
profile and is the smallest contribution.  Lastly, the dashed curve represents
remaining effects such as surface contributions, measurement errors on the
frequencies and effects not included in Eq.~(\ref{eq:error}), such as non-linear
effects not accounted for in Eq.~(\ref{eq:struct_problem}).  Of these three, the
non-linear effects are likely to be the strongest because the data is error free
(except for numerical round-off errors) and no attempt was made to simulate
surface effects in Model A.  As can be seen, the non-linear effects are not
negligible compared to the other sources of error.  This is not entirely
surprising given that the only criteria used for selecting the models
from the grid was the $(\Teff,\logg)$ error box, and consequently shows
the limits of Eq.~(\ref{eq:struct_problem}).  It must also be noted that the
errors in Fig.~\ref{fig:error} tend to compensate each other.  However, one
cannot always expect such fortuitous cancellation to arise in other situations.

\begin{figure}[htbp]
\includegraphics[width=\mycolumn]{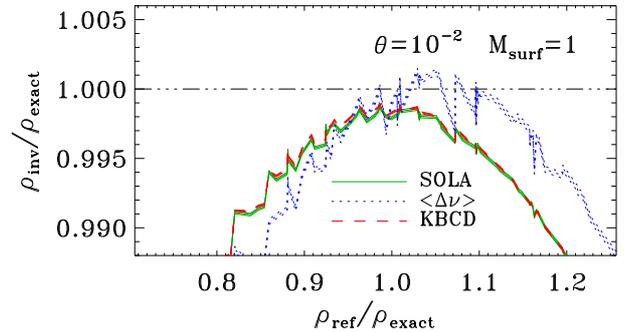} 
\caption{Same as Fig.~\ref{fig:ModelA} \textit{(right, middle panel)}, except
that the non-linear extension described in Sect.~\ref{sect:pre_scaling}
has not been applied to the inversion procedures.
\label{fig:ModelA_unscaled}}
\end{figure}

Figure~\ref{fig:ModelA_unscaled} shows what happens if the non-linear
extension described in Sect.~\ref{sect:pre_scaling} is not applied to the
inversion procedure.  As expected, inversion results are substantially worse
when the mean density of the reference is far from the true mean density.

\begin{figure}[htbp]
\includegraphics[width=\mycolumn]{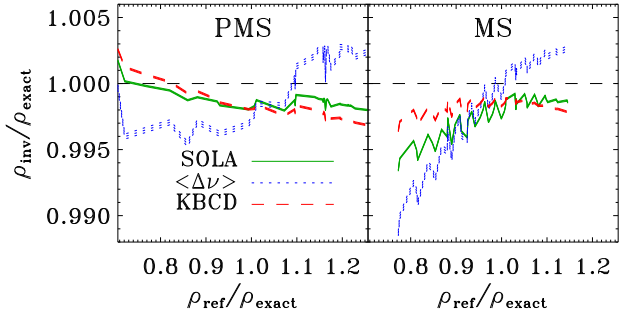} 
\caption{Same as Fig.~\ref{fig:ModelA} \textit{(left, middle panel)},
except that the pre-main-sequence models \textit{(left panel)} and the
main-sequence models \textit{(right panel)} have been separated.
\label{fig:ModelA_separated}}
\end{figure}

Finally, the wiggles observed in Fig.~\ref{fig:ModelA}, especially for
reference models with a lower mean densities, can be partially explained by the
fact that the grid contains a mixture of pre-main-sequence and main-sequence
models.  Indeed, when the two types of models are separated as in
Fig.~\ref{fig:ModelA_separated}, two separate trends can be observed which
likely reflect their differences in structure.

\subsection[Near-surface effects (Model A')]{Near-surface effects (Model \Aprime)}

Model \Aprime\ is nearly identical to Model A, except for an ad-hoc $50\,\%$
increase of the density in the outer $1\,\%$ of the model, the transition being
represented by a hyperbolic tangent function.  The pressure was then
recalculated using the hydrostatic equation, thus resulting in a very slight
offset from the original profile.  The difference in mean density between the
two models is negligible, but the frequencies are noticeably different (see
Fig.~\ref{fig:echelle1}) due to the fact that sound waves spend more time in the
outer layers of a star.

\begin{figure*}[htbp]
\includegraphics[width=\textwidth]{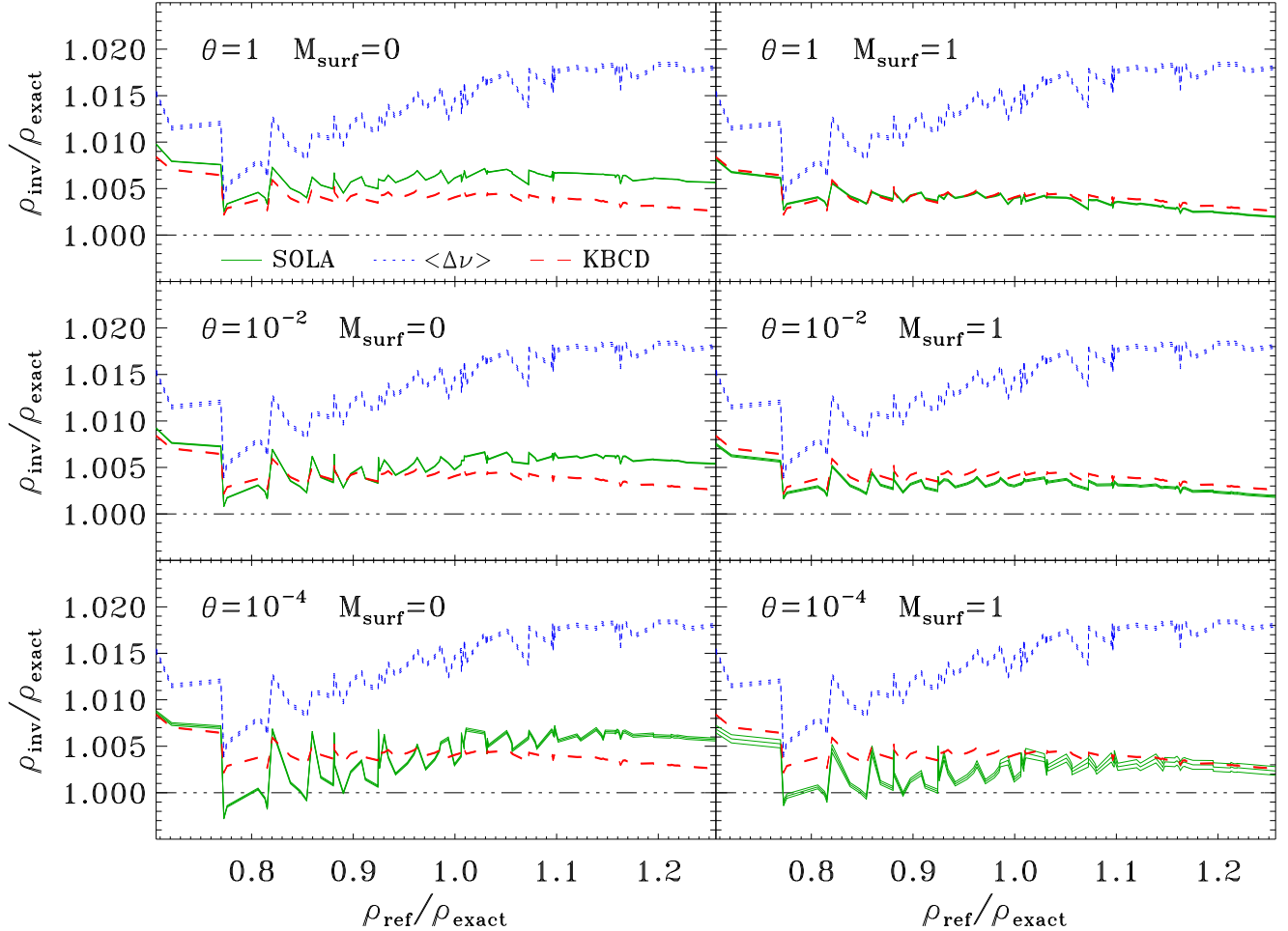} 
\caption{Same as Fig.~\ref{fig:ModelA} except for Model \Aprime.
\label{fig:ModelA_bis}}
\end{figure*}

Figure~\ref{fig:ModelA_bis} shows the inversion results for this model, using
the same set of modes as in the previous case.  Overall, the results are
worse than in the previous case.  Furthermore, the scaling law based on the
large frequency separation yields results which are substantially offset from
the true mean density, when compared to the other procedures.  This could be
expected given the fact that the large frequency separation, which
asymptotically gives the travel time of a sound wave from centre to surface, is
very sensitive to surface conditions.  This high sensitivity to surface
conditions is also reflected in the associated averaging kernels which tend to
have large amplitudes near the surface.  The SOLA inversions without surface
corrections also seem to be affected, but to a lesser degree.  In particular, it
produces worse results than the KBCD approach for the higher density
models, which is the opposite from the previous case.  If surface corrections
are included in the SOLA approach, then these produce slightly better results
than the KBCD approach.  Once more, reducing $\theta$ does not improve
the SOLA inversions but only serves to increase the $1\sigma$ error bar.

\subsection{Uncertainties in the stellar physics (Model B)}

Model B is different in a number of ways from Model A and also from the models
in the grid.  It has a different chemical composition and includes the effects
of diffusion and rotational mixing, thereby providing a way to test the effects
of unknown physics on mean density inversions.  Specifically, the chemical
abundances are $(X,Y,Z) = (0.735,0.250,0.015)$ and follow the more recent solar
mixture given in \citet{Asplund2005}.

\begin{figure}[htbp]
\includegraphics[width=\mycolumn]{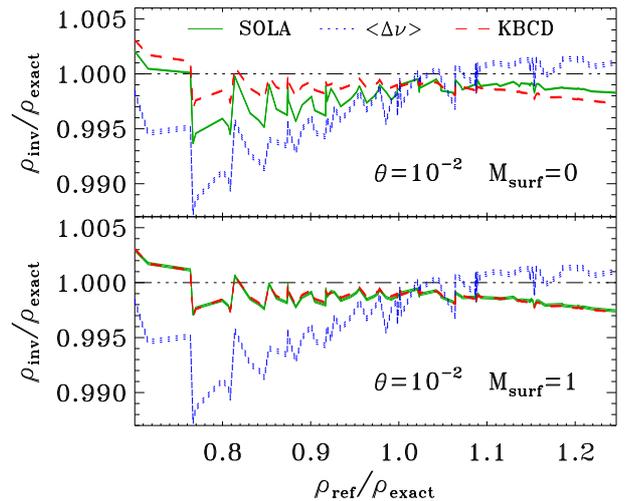} 
\caption{Same as Fig.~\ref{fig:ModelA} except for Model B.
\label{fig:ModelB}}
\end{figure}

The results are shown in Fig.~\ref{fig:ModelB}, using once more the same set of
modes.  These results are in fact similar to those obtained for Model A and shows
that all of the inversion procedures presented here seem to be fairly robust
to unknown physics.

\subsection{The effects of mode misidentification}

\begin{figure*}[htbp]
\includegraphics[width=\textwidth]{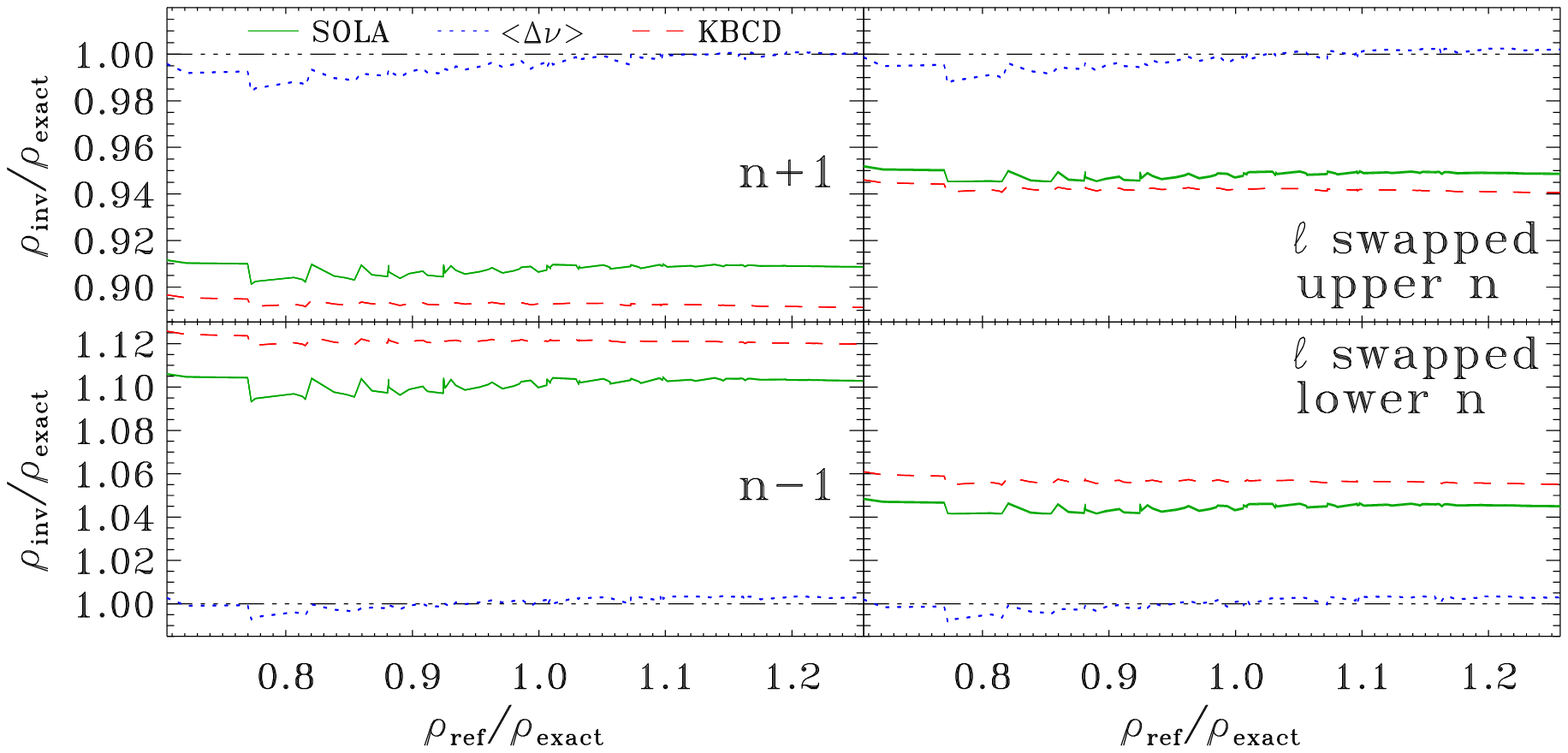} 
\caption{Inversion results for Model A, using the same modes as in
Fig.~\ref{fig:ModelA} but with various erroneous mode identifications, as
indicated in the panels.  In all of the panels, $\beta = 10^{-8}$, $\theta =
10^{-4}$ and $\Msurf = 0$.
\label{fig:misidentification}}
\end{figure*}

We then investigate the effects of misidentifying the radial orders of the
modes.  Figure~\ref{fig:misidentification} (left panels) shows mean density
estimates for Model A, using the same set of observed modes as previously, but
identified as $n=16-26$ (upper panel) and $n=14-24$ (lower panel).  As can be
seen in these figures, the scaling law based on the large frequency separation
provides the best results, whereas the other methods are off by $10 \%$.  A
simple explanation is that the relative change over one radial order of
$\Delta\nu$ is smaller than that of $\nu$, given that $\Delta\nu$ goes to a
constant when $n$ goes to infinity, whereas $\nu$ is roughly proportional to
$n$.

Another way of misidentifying modes is by labelling the $\l=0$ modes as $\l=1$
and vice-versa \citep[\eg][]{Benomar2009a}.  This, however, also leads to
detecting a different set of $\l=2$ frequencies given that these will always be
close to the $\l=0$ frequencies.  In order to investigate this scenario, we swap
the $\l=0$ and $1$ identifications and create a fictitious set of $\l=2$
frequencies in such a way as to preserve the original small frequency
separation, averaged over adjacent radial orders.  Two possible ways of
assigning radial orders exist and are given in
Table~\ref{tab:l_misidentification}. The corresponding mean density estimates
are given in the right panels of Fig.~\ref{fig:misidentification}.  Once more,
the large frequency separation gives the best results, the other methods being
off by about $5 \%$, \ie\ about half the amount as in the previous case when
only $n$ was misidentified.  One way of interpreting this is to view this as a
$\pm 1/2$ offset on the radial order, and then to apply the same argument as
previously.

\begin{table}[htbp]
\caption{Different ways of assigning radial orders when misidentifying the $\l$ value.
\label{tab:l_misidentification}}
\begin{tabular}{ccc}
\hline
\hline
\textbf{True id.} &
\textbf{Upper} $n$ &
\textbf{Lower} $n$ \\
\hline
$\l=1$, $n=15-25$ &
$\l=0$, $n=16-26$ &
$\l=0$, $n=15-25$ \\
$\l=0$, $n=15-25$ &
$\l=1$, $n=15-25$ &
$\l=1$, $n=14-24$ \\
Fictitious  &
$\l=2$, $n=15-25$ &
$\l=2$, $n=14-24$ \\
\hline
\end{tabular}
\end{table}

These observations provide a way of identifying modes: one can calculate the
mean density using the large frequency separation and another method such as the
KBCD approach or a SOLA inversion and search for the values of $\l$ and $n$ for
which the two methods agree to within $1 \%$ or so.  This procedure, although
different from the methods described in \citet{Bedding2010} and
\citet{White2011}, is based on the same principles.  Indeed, the method
presented here and the approach from \citet{Bedding2010} both amount to scaling
a model (or a comparison star) so that the large frequency separation agrees
with that of the observed star, then adjusting the identification so that the
offset between the two sets of frequencies is minimised.  \citet{White2011}
achieves the same result by directly comparing their model-based values for the
large frequency separation and $\epsilon$, the additive offset from the
asymptotic formula for p-mode frequencies, to those from observed stars.  Of
course, surface effects acting upon the large frequency separation may lead to
an offset even for the correct identification, although we expect such an offset
to be smaller than what is shown in Fig.~\ref{fig:ModelA_bis}, which was
calculated for a rather drastic modification of the surface.  \citet{White2011}
have nonetheless pointed out situations where $\epsilon$ can vary rapidly
thereby making mode identification more difficult.

\subsection{Kernel mismatch}
\label{sect:kernel_mismatch}

\begin{figure}[htbp]
\includegraphics[width=\mycolumn]{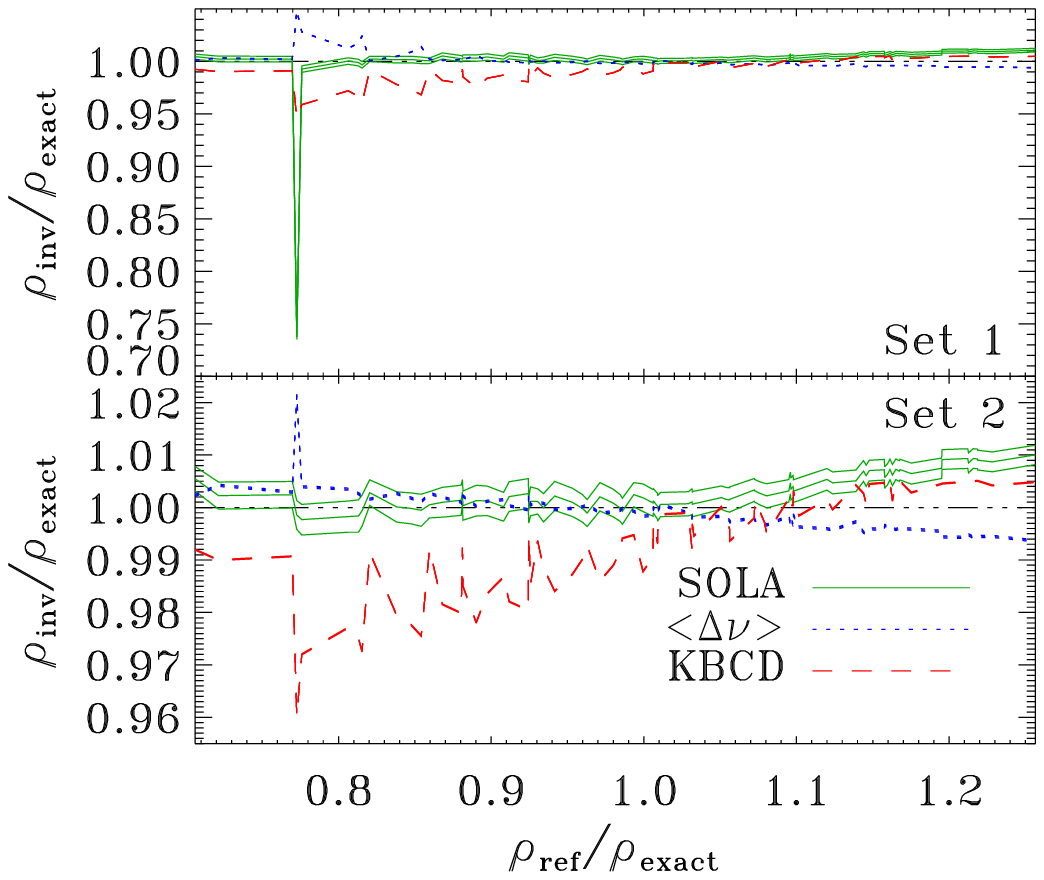} 
\caption{Inversion results for Model A for two similar sets of modes (see text
for details).  The parameters used in these inversions are $\beta=10^{-8}$,
$\theta=10^{-4}$ and $\Msurf = 0$.
\label{fig:non_linear}}
\end{figure}

\begin{figure*}[htbp]
\begin{tabular}{cc}
\includegraphics[width=\mycolumn]{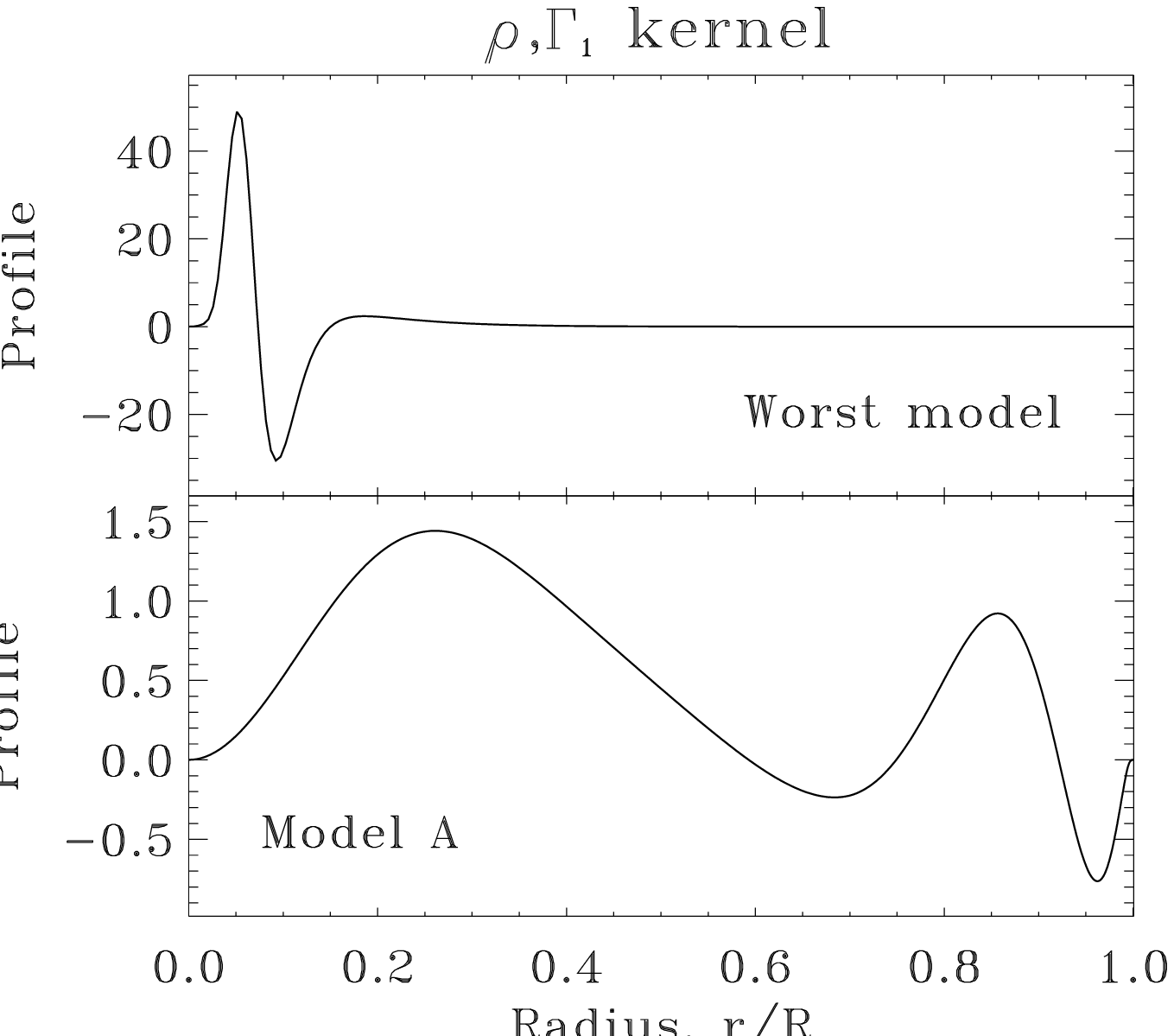} &
\includegraphics[width=\mycolumn]{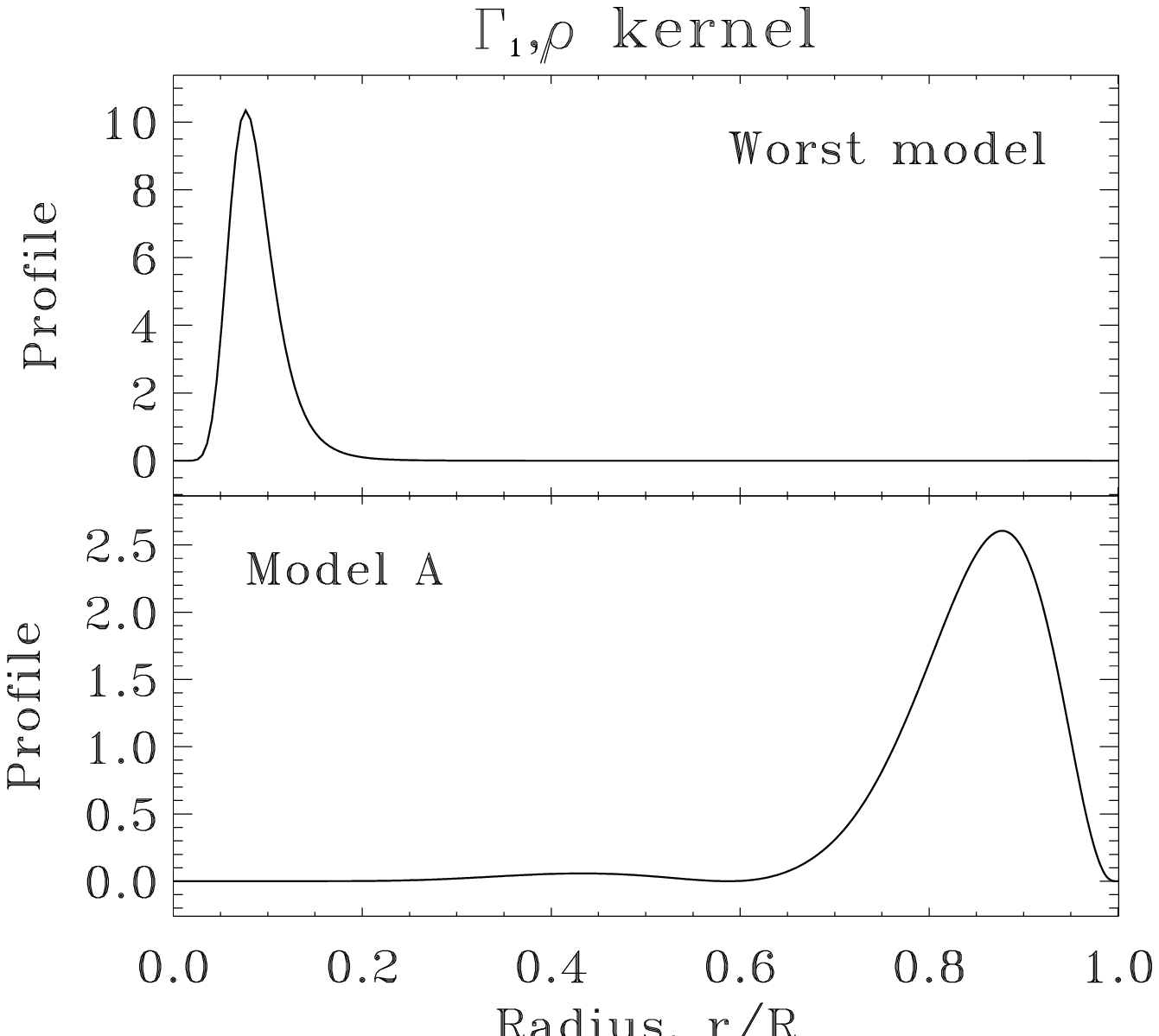} 
\end{tabular}
\caption{Kernels of $n=1$, $\l=4$ mode for Model A (lower panels) and the worst
model (upper panels).
\label{fig:kernels}}
\end{figure*}

It it also interesting to test the various inversion techniques on different sets
of modes.  We therefore consider the following 2 sets:
\begin{itemize}
\item[$\bullet$]\textbf{Set 1}: $n=2-10$ for $\l=0,2,3$, $n=5-10$ for $\l=1$, $n=1-10$ for $\l=4$.
\item[$\bullet$]\textbf{Set 2}: The same as \textbf{Set 1}, without the $n=1$, $\l=4$ mode.
\end{itemize}
The mean density estimates are shown in Fig.~\ref{fig:non_linear}.  As can be
seen by comparing the two figures, including the $n=1$, $\l=4$ mode has a
drastic effect on the SOLA inversion result for one of the models, which we
shall call the ``worst model''.  The explanation for this is quite simple. 
Figure~\ref{fig:kernels} shows the kernels associated with this mode for both
the worst model and for Model A.  Clearly, the kernels in the worst model are
dominated by a g-mode behaviour (even if the eigenfunctions still seem to be
dominated by a p-mode character), whereas those of Model A are typical of
p-modes.  Such a difference on the kernels inevitably leads to a large
discrepancy between the actual frequency shift for that particular mode, and the
frequency shift which would be deduced by applying the linear approximation
given in Eq.~(\ref{eq:struct_problem}).  This, of course, then affects any mean
density estimates, particularly a SOLA inversion which explicitly optimises the
averaging and cross-term kernels using the individual kernels.

\section{Observed stars}
\label{sect:observed_stars}

\subsection[alpha Cen B]{$\alpha$ Cen B}

Having tested the inversion techniques on stellar models in a number of ways, we
now try them out on observed stars.  A particularly interesting example is the
nearby K1 main-sequence star \object{$\alpha$ Cen B}.  Its close proximity
allows precise interferometric measurements of its angular radius
\citep{Kervella2003}.  When combined with a reprocessed Hipparchos
parallax \citep{Soderhjelm1999}, this yields a radius $R = 0.863 \pm
0.005\,R_{\odot}$. Given that \object{$\alpha$ Cen B} is part of a multiple
system, it is possible to determine its mass through orbital parameters.  Hence,
\citet{Pourbaix2002} obtained a mass of $M = 0.934 \pm 0.0061\,M_{\odot}$ using
the parallax from \citet{Soderhjelm1999}.  This then yields a mean density of 
$2.046 \pm 0.058$ $\gcm$.  A time series of radial velocities was
obtained by \citet{Kjeldsen2005} using the Very Large Telescope (VLT)
and the Anglo-Australian Telescope (AAT).  A Fourier analysis of this
time series yielded a set of 37 frequencies with harmonic degrees $\l=0$ to $3$,
and radial orders, $n$, between $17$ and $32$.

\begin{figure}[htbp]
\includegraphics[width=\mycolumn]{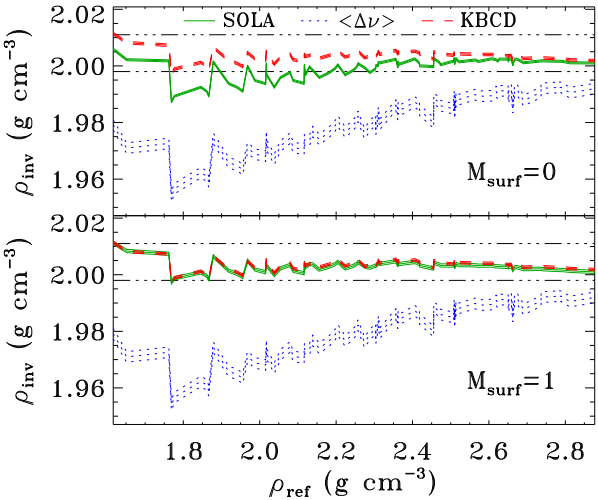}
\caption{Mean density estimates for \object{$\alpha$ Cen B}.  The SOLA
parameters are $\beta = 10^{-8}$ and $\theta=10^{-2}$, and $\Msurf$ is indicated
in the panels.  The vertical triple-dot-dashed line corresponds to the minimum
and maximum mean densities from \citet{Kjeldsen2008}.
\label{fig:alpha_Cen_B}}
\end{figure}

Given that the mass of \object{$\alpha$ Cen B} is close to the mass range
covered by the grid from the previous section, we simply reused the models from
this grid as reference models from which to estimate the mean density of
\object{$\alpha$ Cen B}.  The results are shown in Fig.~\ref{fig:alpha_Cen_B}.
These turn out to be similar to the results obtained in the previous sections. 
The tendencies are the same, as is the spread in inverted mean densities, except
for the results from the $\left< \Delta\nu \right>$ scaling where the spread is
somewhat larger.  Nonetheless, a noticeable difference appears between the
estimates from the $\left< \Delta\nu\right>$ scaling and the two other methods. 
This suggests that surface effects play an important role in this star.

It is then interesting to see how these densities compare with previous
studies.  Table~\ref{tab:alphaCenB} gives a summary of previous and
current estimates of the mean density of \object{$\alpha$ Cen B}.  As was
mentioned above, the mass from \citet{Pourbaix2002} and the radius from
\citet{Kervella2003} lead to a mean density of $2.046 \pm 0.058$ $\gcm$. The
results from the KBCD and SOLA approaches are $2$ to $3\,\%$ lower than this
value, but remain within the error bars.  They agree with the range of
values, delimited by the triple-dot-dashed horizontal lines in
Fig.~\ref{fig:alpha_Cen_B}, obtained in \citet{Kjeldsen2008} where the KBCD
approach is developed.  The seismic study in \citet{Eggenberger2004} leads to a
mean density of 1.998 $\gcm$ which also falls within this range, although they
did increase the error bar on the radius to reach this value (the error bar on
the mass seems to have been ignored in their study).  On the other hand, the
$\left< \Delta\nu \right>$ scaling leads to values below all of these
previous studies and which mostly lie outside the error bars of the
non-seismic mean density.  Hence, using the KBCD or SOLA approaches leads to
better agreement with the non-seismic constraints and tends to confirm the idea
that the $\left<\Delta\nu\right>$ scaling is affected by surface effects.

\begin{table}[htbp]
\caption{Mean density of \object{$\alpha$ Cen B}.
\label{tab:alphaCenB}}
\begin{center}
\begin{tabular}{lccc}
\hline
\hline
\textbf{Source} & $\bar{\rho}$ \\
& $\left(\gcm\right)$ \\
\hline
Non-seismic\tablefootmark{a} & $1.988 - 2.104$ \\
\citet{Eggenberger2004} & $1.998$ \\
\citet{Kjeldsen2008}    & $1.998 - 2.011$ \\
SOLA ($\Msurf=0$)\tablefootmark{b} & $1.987 - 2.006$ \\
SOLA ($\Msurf=1$)\tablefootmark{b} & $1.997 - 2.012$ \\
$\left< \Delta \nu \right>$ scaling\tablefootmark{b} & $1.952 - 1.994$ \\
% KBCD\tablefootmark{b} & $1.997 - 2.012$ \\
\hline
\end{tabular} \\
\tablefoottext{a}{Based on the parallax from \citet{Soderhjelm1999}, the interferometric observations
from \citet{Kervella2003} and the orbital parameters from \citet{Pourbaix2002}.}
\tablefoottext{b}{This study.}
\end{center}
\end{table}

\subsection{HD 49933}

We now turn our attention to stars observed by CoRoT.  A particularly
interesting example is the F5 main-sequence star \object{HD 49933} which has
been observed in two separate CoRoT runs, thereby yielding a rich
frequency spectrum.  In what follows, we will use the set of $51$ frequencies
and associated error bars and identifications given in \citep{Benomar2009b},
which takes into account the two observational runs.  An alternative
identification had been favoured before data from the second observational run
was available \citep[\eg]{Appourchaux2008}.  The grid of models used in the
previous sections is not suitable for this star. We therefore downloaded a set
of $42$ pre-main-sequence and main-sequence models from the CoRoT-ESTA-HELAS
website, which lie in the error box $\Teff = 6665 \pm 215$ K and
$\log(L/L_{\odot}) = 0.54\pm 0.02$ dex, based on the values quoted in
\citet{Benomar2010} and references therein.  It must be noted that no attention
was paid to possible differences in chemical composition between these models
and \object{HD 49933}.

\begin{figure}[htbp]
\includegraphics[width=\mycolumn]{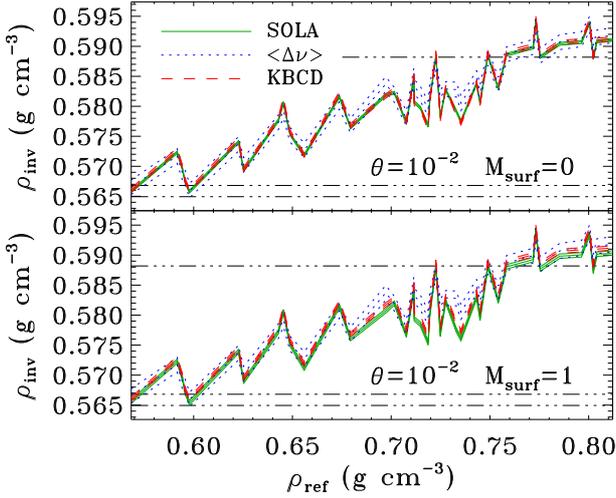}
\caption{Mean density estimates for \object{HD 49933}.  The SOLA parameters are
$\beta=10^{-8}$ and $\theta=10^{-2}$.  The horizontal triple-dot-dashed lines
represent the solutions from \citet{Creevey2011} and \citet{Kallinger2010}, in
the same order as in Table~\ref{tab:HD49933}.
\label{fig:HD49933}}
\end{figure}

The mean density estimates are given in Fig.~\ref{fig:HD49933}.  They fall in
the interval $0.580\pm0.015$ $\gcm$ which corresponds to a $\pm 2.6\%$
range around the mid-value.  This interval is on the lower end of the range of
values covered by the reference (or input) mean densities.  Furthermore, the
estimated mean density slightly increases with the reference mean density, in
much the same way as in the previous examples except on a larger scale.  If we
limit ourselves to reference models with a closer mean density to the results,
this would favour the lower values, but this reasoning must be taken with some
caution since the inversions use the non-linear extension and are
therefore insensitive to the mean density of the input models.  Various other
groups have performed detailed asteroseismic interpretation of his star and have
come up with best fitting models using a variety of techniques.  Here, we will
only focus on studies for which both the mass and radius of the best fitting
model are provided.  \citet{Kallinger2010} searched for a best fitting model in
a dense grid using a $\chi^2$ criteria, whereas \citet{Creevey2011} found two
solutions using both a Levenberg-Marquardt and a Markov Chain Monte Carlo
approach.  Their results are summarised in Table~\ref{tab:HD49933} and
represented as triple-dot-dashed horizontal lines in Fig.~\ref{fig:HD49933}. 
The first solution uses a mode identification in which the $\l$ values are
swapped.  Its mean density is in the upper half of the range of values found
here. The two other solutions use the same identification as the one used here. 
They have very similar mean densities in spite of having different radii and
masses, and are at the lower end of the interval of solutions found here. 
Although further from our mid-value, these solutions are located where the mean
densities are least modified by the inversions.

\begin{table}[htbp]
\caption{Parameters of best fitting models for \object{HD49933} from other studies.
\label{tab:HD49933}}
\begin{tabular}{lccc}
\hline
\hline
\textbf{Source} &
$M$ & $R$ & $\bar{\rho}$ \\
& $\left(M_{\odot}\right)$
& $\left(R_{\odot}\right)$
& $\left(\gcm\right)$ \\
\hline
\citet{Creevey2011}, 1\tablefootmark{a} & 1.12  & 1.39 & 0.588 \\
\citet{Creevey2011}, 2     & 1.20  & 1.44 & 0.567 \\
\citet{Kallinger2010}      & 1.325 & 1.49 & 0.565 \\
\hline
\end{tabular} \\
\tablefoottext{a}{This solution is based on a swapped $\l$ identification.}
\end{table}

\begin{figure}[htbp]
\includegraphics[width=\mycolumn]{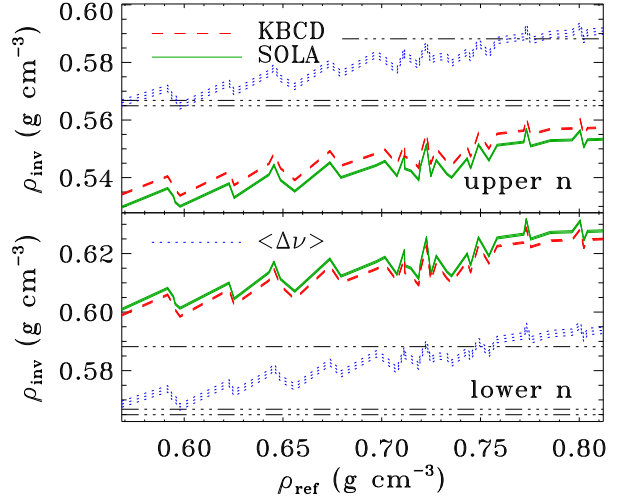}
\caption{Mean density estimates for \object{HD 49933} using swapped $\l$
identifications.  The SOLA parameters are $\beta=10^{-8}$, $\theta=10^{-2}$ and
$\Msurf=1$.  The solutions from \citet{Creevey2011} and \citet{Kallinger2010}
are represented as triple-dot-dashed horizontal lines.
\label{fig:HD49933_identification}}
\end{figure}

Given the difficulties in labelling the modes prior to the second CoRoT run, we
tested out the swapped $\l$ identification, using only the $\l=0$ and $1$
modes.  The results are shown in Fig.~\ref{fig:HD49933_identification} and are
very similar to the results presented in Fig.~\ref{fig:misidentification}. 
Clearly, the models used in this study favour the currently accepted
identification.  In particular, this agrees with the identification favoured by
\citet{Kallinger2010} and with the results from \citet{Bedding2010} who
compared \object{HD 49933} to other observed stars in which the identification is known.

Finally, it is worth noting that surface effects have not manifested themselves
in the results presented here: all three methods yield similar results for the
correct identification.  This suggests that surface effects are less important
in this star compared to other effects deeper within the star. 
Interestingly, \citet{Bedding2010b} recently reported on an
asteroseismic investigation of Procyon, also of spectral type F5.  They showed
that the KBCD approach with a solar exponent, $b=4.9$, did not provide a good
description of the differences between the frequencies of various best fitting
models and those of the star.  Instead, these were better described as a
constant offset, which corresponds to $b=0$.  Given that these differences do
not go to zero as the frequency is reduced, they also interpreted this as
signifying that the structural differences between their best fitting models and
Procyon extend beyond the surface layers.

\subsection{HD 49385}

The star \object{HD 49385} is an evolved G0 type star which was observed by
CoRoT for 137 days during its second long run.  \citet{Deheuvels2010} were able
to detect $27$ $\l=0-2$ frequencies as well as what seems to be mixed modes and
some unconfirmed $\l=3$ frequencies.  In what follows, we will not use the
$\l=3$ modes given their low signal to noise ratio, and will postpone
dealing with the mixed modes to the end of this section, since these
vary rapidly from one model to another, thereby limiting the validity of
Eq.~(\ref{eq:struct_problem}).  A new set of 18 models within the
$(\Teff,\logg) =  (6095 \pm 65 K,\,4.00\pm0.06)$ error box \citep{Deheuvels2010}
was downloaded from the CoRoT-ESTA-HELAS website.  Given the position of this
box in the HR diagram, these models correspond to either pre or
post-main-sequence stars\footnote{As was argued in \citet{Deheuvels2011}, the
presence of mixed modes shows that this is a post-main-sequence star.  Hence,
one would want to limit the inversions to this type of star.  However, for the
sake of comparison, we will also use pre-main-sequence stars and see what impact
this has on the inversions.}.

\begin{figure}[htbp]
\includegraphics[width=\mycolumn]{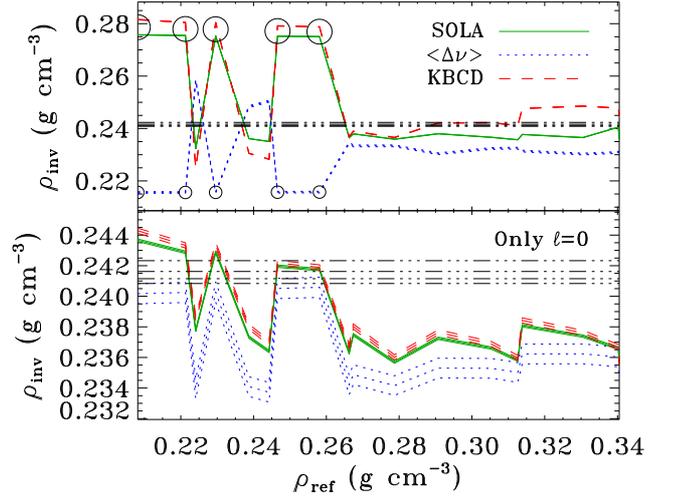}
\caption{Mean density estimates for \object{HD 49385}.  The SOLA parameters are
$\beta = 10^{-8}$, $\theta=10^{-2}$ and $\Msurf=0$.  The horizontal
triple-dot-dashed lines correspond to the best fitting models from
\citet{Deheuvels2011}.  The circled results \textit{(upper panel)} were
obtained for pre-main-sequence models (for clarity, these are not circled in the
lower panel).
\label{fig:HD49385}}
\end{figure}

The mean density estimates are given in Fig.~\ref{fig:HD49385}.  The
horizontal triple-dot-dashed lines are the mean densities corresponding to the four 
families of solutions in \citet{Deheuvels2011}.  In the upper panel, the
results are particularly poor.  This failure is caused by differences in mode
labelling between the pre and post-main-sequence models due to the presence of
mixed modes in the latter.  The inversion results were based on an
identification which corresponds to some of the best-fitting models from
\citet{Deheuvels2011} who applied Newton's method to a grid of models
specifically designed to handle mixed modes.  Such an identification is
ill-adapted for pre-main-sequence models and to a lesser extent
for some of the post-main-sequence models, due to the jumps it introduces in
the $n$ values.  Consequently, the results from the five pre-main-sequence
models, indicated by the circles, stand out as being the worst.  In order to
circumvent these difficulties, a second inversion was carried out, only using
the $\l=0$ modes. This time, the results are far better and are within $2.5 \%$
of the mean densities from \citet{Deheuvels2011} for the SOLA inversions
and for the KBCD approach, and $3.5 \%$ for the scaling law based on the
large frequency separation.  Paradoxically, the pre-main-sequence models
now give the results which are closest to \citet{Deheuvels2011}.  Once
more, the scaling law is somewhat different than the two other approaches which
is probably indicative of surface effects.

In order to investigate the influence of mixed modes on SOLA mean density
inversions, we use a low overshoot model from \citet{Deheuvels2011} with
the \citet{Asplund2005} mixture. Inversions were carried out for three
different sets of modes:
\begin{itemize}
\item[$\bullet$] \textbf{Set 1}: 26 p-modes with $\l=0-2$
\item[$\bullet$] \textbf{Set 2}: Set 1 + the $(n=14,\l=1)$ p-mode
\item[$\bullet$] \textbf{Set 3}: Set 1 + the $(n=10,\l=1)$ mixed
mode\footnote{We note that there there is an offset of 1 between the
radial orders used here and the ones used in \citet{Deheuvels2011}.  This is
merely a difference in mode labelling and consequently does not affect inversion
results.}
\end{itemize}
Two different sets of SOLA parameters were used:
\begin{itemize}
\item[$\bullet$] \textbf{Case 1}: $\beta = 10^{-8}$, $\theta=10^{-2}$ and $\Msurf=1$.  These
      parameters were chosen in accordance with typical inversion parameters
      (the number of modes was sufficient to allow $\Msurf=1$).
\item[$\bullet$] \textbf{Case 2}: $\beta = 0$, $\theta = 0$, $\Msurf = 0$.  These parameters
      correspond to optimising $\Kavg$ only.
\end{itemize}

The results are summarised in Table~\ref{tab:HD49385_mixed_modes}.  As
can be seen in case 1, adding a p-mode or a mixed mode has a similar impact on
both the averaging kernel and the cross-term kernel.  The reduction of the
statistical error on the inverted mean density, $\sigma_{\rhobar}$, is larger
for the mixed mode than for the p-mode, indicating that the mixed mode may be
yielding more information on the mean density.  This seems to be confirmed in
case 2, where only the averaging kernel is being optimised. Indeed, the
averaging kernel including the mixed mode is closer to the target function than
the averaging kernel which uses the extra p-mode.  However, the extra
benefit from the mixed mode remains small and is probably of little use because
the domain for which Eq.~\ref{eq:struct_problem} is valid is likely to be much
smaller for these modes than for ordinary p-modes.  In some cases, non-linear
effects on such modes can even cause a substantial error on the mean density as
was shown in Sect.~\ref{sect:kernel_mismatch}.

\begin{table}[htbp]
\caption{SOLA inversion results for \object{HD 49385}.
\label{tab:HD49385_mixed_modes}}
\begin{center}
\begin{tabular}{lccccc}
\hline
\hline
\textbf{Mode set}
& $\rhobar_{\mathrm{inv}}$ 
& $\sigma_{\rhobar}$
& $\|\Delta\Kavg\|_2$
& $\|\Kcross\|_2$ \\
& ($\gcm$)
& ($\gcm$) & & \\
\hline
\multicolumn{5}{c}{\textbf{Case 1}: $\beta = 10^{-8}$, $\theta=10^{-2}$, $\Msurf=1$} \\
\hline
Set 1 & 0.2415 &  1.63$\times 10^{-4}$ &  0.9788 &  2.261 \\
Set 2 & 0.2415 &  1.61$\times 10^{-4}$ &  0.9737 &  2.254 \\
Set 3 & 0.2415 &  1.46$\times 10^{-4}$ &  0.9753 &  2.247 \\
%Set 4 & 0.2415 &  1.52$\times 10^{-04}$ &  0.9698 &  2.240 \\
\hline
\multicolumn{5}{c}{\textbf{Case 2}: $\beta = 0$, $\theta=0$, $\Msurf=0$} \\
\hline
Set 1 & 0.1932 &  0.104 & 0.4578 &  4.253 \\
Set 2 & 0.3439 &  0.433 & 0.4551 &  6.055 \\
Set 3 & 0.2029 &  0.134 & 0.4495 &  5.224 \\
%Set 4 & 0.3400 &  0.454 & 0.4473 &  6.757 \\
\hline
\end{tabular}
\end{center}
\end{table}

Does this mean that mixed modes are therefore not a useful diagnostic of
core conditions, and that previous studies somehow exaggerated their importance? 
Certainly not.  Rather, it means that the methods presented here are not so good
at exploiting the full diagnostic potential of such modes. Other methods, such
as the one presented in \citet{Deheuvels2011}, or inversions localised
around the stellar core, are needed to fully benefit from these modes.  A
possible explanation for this failure can be described as follows.  Structural
kernels associated with mixed modes and g-modes, such as the ones illustrated in
the upper panels of Fig.~\ref{fig:kernels}, have large oscillatory behaviour
near the centre of the star, which does not match the target function, $T$,
defined Eq.~(\ref{eq:target}).  As a result, their contribution to SOLA
inversions tends to be reduced, as reflected, to some extent, in their inversion
coefficients. On the other hand, their contribution is not reduced in the KBCD
approach or the $\left<\Delta\nu\right>$ scaling law, and this consequently
leads to averaging kernels with highly oscillatory behaviour near the centre
which can also be a source of error.

\section{Discussion}

In this paper we have investigated different techniques for estimating directly
the mean density of a star.  These include the usual scaling law based on the
large frequency separation, an approach described in \citet{Kjeldsen2008}
(the ``KBCD approach''), and a SOLA inversion technique with a
non-linear extension.  We have shown how all three techniques can be put in a
similar form, which allows the construction of associated averaging and
cross-term kernels, useful for characterising what is actually being measured
and for quantifying the error.

A comparison of the three techniques, applied to the \object{sun} and to various
test cases, shows that the KBCD approach and SOLA inversions yield similar
results, whereas the scaling law based on the large frequency separation
produces sub-optimal results.  The averaging kernels produced by the KBCD
approach can come close in quality to those produced by a SOLA inversion and are
substantially better than the averaging kernels associated with the simple
scaling law based on $\left<\Delta\nu\right>$, thereby explaining its relative
success.  Hence, if one only wants to estimate a mean density and does not need
a full diagnostic on the accuracy of the result, the KBCD approach should be
sufficient.  However, for a more detailed analysis, the SOLA method seems ideal
as it is specifically designed to optimise the averaging kernel while reducing
various sources of error.  Furthermore, possible improvements to the SOLA
method, such as introducing weight functions in the cost function
Eq.~(\ref{eq:SOLA}) \citep[\eg][]{Rabello-Soares1999}, may help to fine-tune
this method and should therefore be explored.

As expected from asymptotic theory, near-surface effects are shown to have a
stronger impact on the $\left<\Delta\nu\right>$ scaling law than on the two
other methods, even when surface corrections are not included in the SOLA
approach. However, these two methods require a correct identification of the
modes whereas the $\left<\Delta\nu\right>$ scaling law is much more tolerant of
mode misidentification.  Hence, looking for an $\l$ and $n$ assignment which
provides the best agreement between the simple scaling law and one of the two
other methods can be a useful way of identifying modes.  This approach to mode
identification, although different from the methods presented in
\citet{Bedding2010} and \citet{White2011}, is also based on the same
principles.

Quantitatively, a precision around $0.5\,\%$ or better was reached for the
\object{sun}, for \object{$\alpha$ Cen B} (as based on the spread of the
results), and for various test cases using SOLA inversions (with $\beta=10^{-8}$
and $\theta=10^{-2}$) and the KBCD approach, whereas the scaling law was less
accurate with a precision of $1\,\%$ or worse.  However, when applied to two
stars observed by CoRoT, the spread in values was larger, up to $\pm2.5\,\%$ for
the first two methods, and $\pm 3.5\,\%$ for the $\left<\Delta\nu\right>$
scaling law.  The reason for this remains an open question, but does not seem to
stem from the fact that these are observed stars, as indicated by
further tests involving models instead (results not included) and the
results obtained for \object{$\alpha$ Cen B}. Rather, it may be that in the
part of the HR diagram which contains these stars, there are larger variations
in the density profile with stellar evolution.  A $\pm 3.5\,\%$ spread
is comparable to the spread obtained in \citet{White2011} for stars
with $\Teff$ below $6700$ K.  Above this temperature, \citet{White2011} 
obtained larger differences, which in some cases reached $10\,\%$,
between the true density and that obtained from the $\left<\Delta\nu\right>$
scaling law.  This raises the question as to how accurate the KBCD and SOLA
approaches would be in a similar situation, although one also has to bear in
mind that this temperature range is less relevant for solar-like oscillations. One
notable difference between the results described here and those presented in
\citet{White2011} is that the latter are calibrated to solar values
rather than nearby reference models, which is likely to be an additional source
of error, especially for stars with a large temperature difference with respect
to the sun.

If combined with a $2\,\%$ error bar on measured radii, as is expected
from GAIA, a $2.5\,\%$ to $3.5\,\%$ uncertainty on the mean density
yields an $8.5-9.5\,\%$ error bar on the mass of the star, where the dominant
contribution to the error comes from the uncertainty on the radius. 
Such an error bar will then directly translate itself into a comparable
uncertainty on the inferred mass of any potential exoplanet \citep[\eg][and
references therein]{Guillot2011}.  Accordingly, one may try to obtain more
accurate results through more complicated approaches such as searching for best
fitting models in a grid or applying non-linear inversion techniques.  Such
approaches will overcome the non-linear effects shown Fig.~\ref{fig:error} and
which are beyond the control of the methods presented here.  However, given the
difficulties in optimising the averaging kernel beyond a certain point for a
given set of modes, one is still left with the question as to how much
information is intrinsically available in the pulsation frequencies and how much
is being assumed in these more sophisticated methods.

Finally, it was shown that the methods presented here do not fully
exploit the diagnostic potential of mixed modes.  Indeed, mixed modes have
kernels with highly oscillatory behaviour near the centre which strongly differs
from the target function, $T$, defined in Eq.~(\ref{eq:target}). This reduces
the contribution of mixed modes to SOLA inversions and leads to averaging
kernels with an oscillatory behaviour near the centre for the
$\left<\Delta\nu\right>$ scaling law and the KBCD approach.  Also, the linear
domain for these modes is smaller and can lead to substantial errors when
outside this domain, as shown in Sect.~\ref{sect:kernel_mismatch}.  However,
one may expect these modes to be very useful when applying localised inversion
methods to models which are close to the true stellar structure.

\begin{acknowledgements}
We thank the referee for helpful comments and suggestions which
have improved the manuscript.  We also thank I. Roxburgh, S. Vorontsov and S.
Basu for interesting discussions and answers to questions.  DRR is supported by
the  CNES (``Centre National d'Etudes Spatiales'') through a postdoctoral
fellowship.  This article made use of InversionKit, an inversion software
developed in the context of the European Helio- and Asteroseismology Network
(HELAS), which is funded by the European Commission's Sixth Framework
Programme.  NCAR is sponsored by the National Science Foundation.
\end{acknowledgements}

%%%%%%%%%%%%%%%%%%%%%%%%%%%%%%%%%%%%%%%%%%%%%%%%%%%%%%%%%%%%%%%%%%%%%%%%%%%%%%%
\bibliographystyle{aa}
\bibliography{biblio}
%%%%%%%%%%%%%%%%%%%%%%%%%%%%%%%%%%%%%%%%%%%%%%%%%%%%%%%%%%%%%%%%%%%%%%%%%%%%%%%
\end{document}